\documentclass[useAMS,usenatbib,usegraphicx]{mn2e}

\usepackage{amssymb}
\usepackage{epsfig}
\usepackage{graphicx}
\usepackage{times}
\usepackage{amsmath}
\usepackage{lscape}
\usepackage{array,multirow}
\usepackage{float}

\title[Evolution of Star Formation in the UDS out to z\,=\,\rm{1.6}]{Evolution of Star
  Formation in the UKIDSS Ultra Deep Survey Field - \MakeUppercase{\romannumeral 1}. Luminosity
  Functions and Cosmic Star Formation Rate out to z\,=\,{1.6}}
\author[Alyssa B. Drake et al.]{Alyssa B. Drake$^{1}$\thanks{E-mail:
    abd@astro.livjm.ac.uk}, Chris Simpson$^{1}$, Chris A. Collins$^{1}$,
  Phil A. James$^{1}$, Ivan K. Baldry$^{1}$, \newauthor Masami
  Ouchi$^{2,3}$, Matt J. Jarvis$^{4,5,6}$, David G. Bonfield$^{5}$, Yoshiaki Ono$^{2}$, Philip N. Best$^{7}$,
  \newauthor Gavin B. Dalton$^{4,8}$, James S. Dunlop$^{7}$,
  Ross J. McLure$^{7}$, Daniel J. B. Smith$^{5}$ 
  \\$^{1}$Astrophysics Research Institute, Liverpool John
 Moores University, Twelve Quays House, Egerton Wharf, Birkenhead,
 CH41 1LD
\\$^{2}$Institute for Cosmic Ray Research, The University of Tokyo,
Kashiwa 277 8582, Japan
\\$^{3}$Kavli Institute for the Physics and Mathematics of the
Universe, WPI, The University of Tokyo, Chiba 277 8583,
Japan 
\\$^{4}$Astrophysics, Department of Physics, Keble Road, Oxford, OX1
3RH
\\$^{5}$Centre for Astrophysics, Science and Technology Research
Institute, University of Hertfordshire, Hatfield, Herts, AL10 9AB
\\$^{6}$Physics Department, University of the Western Cape, Private
Bag X17, Bellville 7535, South Africa
\\$^{7}$SUPA, Institute for Astronomy, Royal Observatory, Blackford
Hill, Edinburgh, EH9 3HJ
\\$^{8}$RALSpace, STFC Rutherford Appleton Laboratory, HSIC, Oxford, OX11 0QX}

\begin{document}               

\pagerange{\pageref{firstpage}--\pageref{lastpage}} \pubyear{2012}

\maketitle

\label{firstpage}

\begin{abstract} 
We present new results on the cosmic star formation history in the
SXDS-UDS field out to $z=1.6$. We compile narrow-band data
from the Subaru Telescope and the Visible and Infrared Survey
Telescope for Astronomy (VISTA) in conjunction with broad-band data
from the SXDS and UDS, to make a selection of 5725
emission-line galaxies in 12 redshift
slices, spanning 10 Gyr of cosmic time. We determine photometric
redshifts for the sample using 11-band photometry, and use a spectroscopically confirmed subset
to fine tune the resultant redshift distribution. We use the maximum-likelihood
technique to determine luminosity functions in each redshift slice and
model the selection effects inherent in any narrow-band selection
statistically, to obviate the retrospective corrections ordinarily
required. The deep
narrow band data are sensitive to very low star formation rates (SFRs), and allow an
accurate evaluation of the faint end slope of the Schechter function,
$\alpha$. We find that $\alpha$ is particularly sensitive to the assumed faintest broadband
magnitude of a galaxy capable of hosting an emission line, and propose
that this limit should be empirically motivated. For this analysis we base our threshold
on the limiting observed equivalent
widths of emission lines in the local Universe. We compute the characteristic SFR of
galaxies in each redshift slice, and the integrated SFR density,
$\rho_{\rm{SFR}}$. We find our results to be in good agreement with the
literature and parametrize the evolution of the SFR density as
$\rho_{\rm{SFR}} \propto (1 + z)^{4.58}$ confirming a steep decline in star formation activity
since $z \sim 1.6$.

\end{abstract}


\section{Introduction}
\label{intro}

A key goal and surmountable challenge for observational cosmology
today is to assemble a reliable picture of the formation and evolution
of galaxies, and the build up of stellar mass in the Universe. The
secular evolution of galaxies is a result of the formation of new stars, and so
understanding the processes triggering and truncating star formation
in addition to placing constraints on the evolution of the star
formation rate (SFR), in turn allows us to constrain theoretical models
of galaxy formation and evolution. A multitude of star formation
indicators have been utilized in evaluating the SFR,
however each of these is subject to its own biases and
caveats, in addition to the limitations technology places on
the redshift range where each method is applicable. 

Ultraviolet (UV) flux provides the most direct measurement of SFR  by tracing hot
young massive stars (e.g. \citealt{Bunker04}, \citealt{Ouchi04},
\citealt{Arnouts05}, \citealt{Baldry05},
\citealt{Bouwens10}) however UV measurements are subject to large dust
corrections and so carry a large uncertainty in addition to the danger
of missing heavily obscured regions of star formation. The dust
enshrouding young stars can itself be used to trace star formation, as
it re-radiates the incident UV radiation in the
infrared (e.g. \citealt{Perez05}, \citealt{LeFloch05}). However, this
method in turn struggles to
detect some \emph{unobscured} star formation. Less commonly used methods
include using radio emission to trace supernova remnants associated
with short-lived massive stars (e.g. \citealt{Mauch07}) and tracing
X-ray emission produced in X-ray
binaries (of which one component is a massive O or B star). Alternatively, emission lines generated in
H{\sc~ii} regions can be used to trace either the recombination of
atoms or the forbidden lines that arise in this low-density environment.
Emission lines in the optical and UV are of course subject to dust
corrections much as many other tracers, however H$\alpha$ luminosity
is particularly well calibrated for extinction effects. 

The evolution of the star formation
rate density, $\rho_{\rm{SFR}}$, is well studied, and a strong upward
evolution is found with redshift out to $z\sim1$ indicating
that the peak of star formation activity occurs at $z>1$. The
compilation of \cite{Hopkins06} provides a broad overview of this
relationship out to $z\sim6$ but demonstrates that, even within the
relatively well-constrained epoch, estimates of SFR evolution still vary
according to indicator used. For instance the evolution of
$\rho_{\rm{SFR}}$ is often parametrized as $\rho_{\rm{SFR}} \propto
(1+z)^{\gamma}$, and in one of the earliest studies (using a rest frame UV
colour to measure SFR) \cite{LillyCFRS96} found $\gamma = 4$. However, their sample was hampered by a
small survey area, and the larger UV-selected samples of
e.g. \cite{CSB99}, \cite{Wilson02} and \cite{Prescott09}  found much shallower
slopes of $\gamma= 1.5, 1.7\pm 1.0$ and $2.5\pm0.3$
respectively. Meanwhile, the
FIR luminosity found in \cite{LeFloch05} translates to a $\gamma = 3.2^{+0.7}_{-0.2}$.

Studies making use of continuum emission
as a measure of the star formation activity of galaxies, whether it be
UV emission, or its re-radiation in the infrared, allow us to quantify star formation in bright galaxies even
down to low SFRs. The samples however are generally small, as sources not only require
spectroscopy but must also be apparent in broad band
filters in order to be detected --- the imposed broad band
flux limit means that samples in the optical become effectively
stellar-mass limited, and therefore do not detect the lower mass
population of galaxies 
responsible for an important contribution to the overall $\rho_{\rm{SFR}}$.

The use of narrow-band filters to detect emission-line galaxies
provides a complementary technique to broad-band-selected samples. A
narrow-band survey selects objects based on the strength of line emission, and so
while the equivalent width of emission lines comes into play, samples
are close to SFR-limited, allowing the
analysis of SFR over a much broader range of stellar masses.
Recently, many authors have utilised this technique to
compile large samples of star-forming galaxies and examine the evolution
of the SFR as traced by either H$\alpha$ luminosity (which
scales directly with SFR; \citealt{Kennicutt92}) or alternatively the [O{\sc~iii}] and
[O{\sc~ii}] forbidden lines which give a more loosely calibrated 
indication of the rate of star formation. 

H$\alpha$ surveys have been hugely successful, from the
pioneering work of \cite{Bunker95} detecting only a handful of emitters,
to the recent advances from the High
Redshift Emission Line Survey (HiZELS; \citealt{Geach08},
\citealt{Sobral09}) that detected thousands of
emission-line sources. Many authors have
taken this approach, and the H$\alpha$ luminosity function is
well-studied in a variety of environments (e.g. \citealt{Gallego95},
\citealt{Yan99}, \citealt{Tresse02},
\citealt{Fujita03}, \citealt{Kodama04}, \citealt{Umeda04},
\citealt{Shioya08} and 
\citealt{Villar08} to name
but a few). Since infrared spectroscopy is required to trace the H$\alpha$ line beyond
$z=0.4$, strong emission lines blueward of H$\alpha$ are required to
study the higher
redshift population using existing optical data sets. The [O{\sc~ii}] doublet
for example has been traced to $z=1.6$ by \cite{Hogg98}, \cite{Hicks02},
\cite{Teplitz03}, \cite{Ajiki06}, \cite{Drozdovsky05}, \cite{Gilbank10} and
\cite{Ly12}.

\cite{Ly12} incorporated the [O{\sc~iii}] tracer as well and derived
luminosity functions in 11 redshift slices in the Subaru Deep Field
based on H$\alpha$, [O{\sc~iii}] and [O{\sc~ii}] luminosities. Fitting
a Schechter function to their data, they
found strong evolution in the faint-end slope, $\alpha$, and the
characteristic number density
$\phi^{*}$, but little evolution in the characteristic luminosity of
galaxies, L$^{*}$. They interpreted these
results as a comparatively 
stronger evolution of the low-luminosity galaxy population than those with
higher intrinsic 
luminosities. They found a steep evolution in $\rho_{\rm{SFR}}$, in good agreement with previous
results, a trend confirmed by the work of \cite{Sobral12} who
compiled the first view of star formation from $z=0.4$ out to
$z=2.2$ using only H$\alpha$-selected galaxies. \cite{Sobral12} parametrized the
evolution in $\rho_{\rm{SFR}}$ as log
$\rho_{\rm{SFR}} = -2.1/(z+1)$ suggesting a $\gamma\sim 3$ across the
redshift range $0 < z < 1.6$. Conversely however, they
found no evolution of $\alpha$, and set the parameter to $-1.6\pm0.08$
for the last 11 Gyr, concluding that previous claims of evolution stemmed
from heterogeneous samples. This value of $\alpha$ was in good agreement with
\cite{Hayes10} who found a slope of $\alpha=-1.72\pm0.20$ at
$z=2.2$ for an H$\alpha$-selected sample. Even consistently
selected samples however (i.e. those using the same emission line
    to derive each luminosity function) still show discrepancies in the value of
$\alpha$ and its evolution with redshift (or lack of). The H$\alpha$ emitters at $z=2.2$ found in
\cite{Tadaki11} for example result in a slope of $\alpha=-1.37$,
much shallower than the studies of \cite{Hayes10} and \cite{Sobral12} and
suggesting little evolution from high redshift to the present day. \cite{Hayes10}
however, while
in agreement with a steep value of $\alpha$ argued that their results did
indeed confirm an evolution in
$\alpha$ compared to the shallower slope of $\alpha= -1.35$ at $z=0$ determined by
\cite{Gallego95}. Evidently, values of $\alpha$ across redshift remain unclear, and tentative
claims of the parameter's evolution divide opinion further still. It
is important therefore to better
understand the factors which most affect the derivation of $\alpha$,
in order to place more meaningful constraints on the faint end slope.

 This paper demonstrates how the maximum likelihood method can be
     used to determine luminosity functions, and the benefits of this
     approach over existing methods. Narrow-band selection is inherently
     sensitive to only a certain region of parameter
     space, and retrospective corrections are ordinarily applied in
     order to account for the undetected objects. Here, we use a
     maximum likelihood analysis to model the selection effects
     statistically and robustly under
     assumptions about the underlying galaxy population. We explain
     in detail how we implement this method, and confirm the technique
   by replicating the results of \cite{Sobral12_double} and \cite{Sobral12}.

Our luminosity functions are derived
from narrow-band-selected emission-line
galaxies in the 
Subaru/XMM-Newton Deep Field (SXDF), coincident with the Ultra Deep Survey
(UDS) field of the UK Infrared
Telescope (UKIRT)'s Infrared Deep Sky Survey (UKIDSS). We use the H$\alpha$,
[O{\sc~iii}] and  [O{\sc~ii}] emission lines to determine field luminosity
functions in 12 redshift slices, spanning
10 Gyr of cosmic time, out to $z\sim1.6$. We place constraints on
the characteristic SFR in each redshift slice, and make estimates of
$\rho_{\rm{SFR}}$ for comparison to values in the literature. The outline of this
paper is as follows. In Section \ref{data}, we summarize the data used
in this analysis, in Section \ref{sample} we discuss the source extraction and
selection of narrow-band emitters, and present the
colour--magnitude plots of narrow-band selected objects. Section
\ref{Redshifts} discusses our spectroscopically-confirmed subset, the star--galaxy separation technique applied in order to
remove late-type stellar contaminants, our photometric redshift analysis
and the resultant photometric redshift distributions, and finally our
method of redshift slice
assignment for emitters selected in each filter. Section \ref{ML}
details our application of the maximum likelihood technique for determining luminosity
functions, and discusses in detail the factors affecting the
narrow-band selection technique. We describe also how this analysis carefully models the
selection effects ordinarily corrected for retrospectively. In
Section \ref{results} we present luminosity functions (LFs) and derive
the characteristic SFR and $\rho_{\rm{SFR}}$ in each
redshift slice for comparison to literature results. Section
\ref{discussion} discusses the factors affecting our results and the
effect of varying assumptions about the underlying galaxy population
on the resultant values of $\alpha$. We summarize our conclusions in Section \ref{conclusions}.

An H$_0$ = 70 kms$^{-1}$ Mpc$^{-1}$, $\Omega_M$ =0.3 and
$\Omega_{\Lambda}$ =0.7 cosmology is assumed throughout, and  all
magnitudes are in the AB system.

\section{Data}
\label{data}

The data used (summarized in Table \ref{table:data}) are taken from the
SXDF-UDS field in order to make use of the deep optical data from the
Subaru/XMM-Newton Deep Survey (SXDS), in addition to deep infrared
imaging from UKIDSS UDS. 

\begin{table*}
\caption{Summary of photometric data in all narrow and broad-band
  filters from Subaru, VISTA, CFHT, UKIRT and Spitzer. }
\label{table:data}
\begin{center}
\begin{tabular}{ccccccc} \hline
Telescope:Instrument & Filter & Filter Type & Central wavelength &
FWHM & 5
sigma limit (2'' aperture) & Citation \\
\hline
Subaru: Suprime-Cam & NB503 & NB &5029\AA & 73\AA & 24.91 & SXDS; \cite{Ouchi08}\\
Subaru: Suprime-Cam & NB570 & NB &5703\AA & 68\AA & 24.52 & SXDS; \cite{Ouchi08} \\
Subaru: Suprime-Cam & NB816 & NB &8150\AA & 119\AA & 25.65 & SXDS; \cite{Ouchi08} \\
Subaru: Suprime-Cam & NB921 & NB &9183\AA & 131\AA & 25.38 & SXDS; \cite{Ouchi09} \\ 
VISTA: VIRCAM & NB980 & NB &9781\AA & 105\AA & 23.17 & Jarvis et al., (in prep) \\
VISTA: VIRCAM & NB990 & NB &9909\AA  & 105\AA & 23.54 & Jarvis et al., (in prep) \\
CFHT: Megacam & \emph{u} & BB &3740\AA & 500\AA & 26.72 & Foucaud et al., (in prep)\\
Subaru: Suprime-Cam & \emph{B} & BB &4473\AA & 1079\AA & 27.39 & \cite{Furusawa08} \\
Subaru: Suprime-Cam & \emph{V} & BB &5482\AA & 984\AA & 27.10 & \cite{Furusawa08}\\
Subaru: Suprime-Cam & \emph{R} & BB &6531\AA & 1160\AA & 26.85 & \cite{Furusawa08} \\
Subaru: Suprime-Cam & \emph{i} & BB &7695\AA & 1543\AA & 26.66 & \cite{Furusawa08} \\
Subaru: Suprime-Cam & \emph{z} & BB &9149\AA & 1384\AA& 25.95 & \cite{Furusawa08}\\
UKIRT: WFCAM & J & BB &12500\AA & 1570\AA& 24.98 & UKIDSS UDS; Almaini et al., (in prep) \\
UKIRT: WFCAM & H & BB &16500\AA & 2910\AA& 24.27 & UKIDSS UDS; Almaini et al., (in prep) \\
UKIRT: WFCAM & K & BB &22000\AA & 3530\AA& 24.59 & UKIDSS UDS; Almaini et al., (in prep) \\
Spitzer: IRAC & ch1 & BB &35500\AA & 7411\AA& 23.55 & Spitzer SpUDS: PI Dunlop \\
Spitzer: IRAC & ch2 & BB &44900\AA & 10072\AA& 22.88 &Spitzer SpUDS: PI Dunlop \\
\hline
\end{tabular}
\end{center}
\end{table*} 

%\citep[UDS;][]{Foucaud06} 
The SXDS (\citealt{Furusawa08}) covers 1.3 square degrees and delivers
deep optical imaging in the \emph{B, V, R, i} and \emph{z} bands in
five overlapping pointings of Suprime-Cam.

UKIDSS, comprised of 5 sub surveys of various areal coverages and depths
\citep{Lawrence07} targeted the SXDF with the deepest of these surveys, the UDS
(\citealt{Foucaud06}, Almaini et al., [in prep]) covering 0.77 sq degrees. Here we utilise
the Eighth Data Release (DR8) from this survey (details in Table \ref{table:data}). Additionally we
incorporate \emph{u} band data across the entire field from Megacam
on the Canada France Hawaii Telescope (CFHT), and
Spitzer IRAC imaging from the Spitzer-UDS (SpUDS) survey in channels 1
and 2.

Narrow-Band imaging was obtained from Suprime-Cam in
four narrow-band filters, and smoothed using a Gaussian kernel to
recover the same point spread function (PSF) as the broad-band optical data
from Subaru (0.82
arcseconds). Details of these data can be
found in \cite{Ouchi08} and \cite{Ouchi09}. 

In addition, we make use of two
narrow-band filters on the Visible and Infrared Survey Telescope for
Astronomy (VISTA). VISTA's visible-infrared camera (VIRCAM;
\citealt{Dalton06}), is a wide-field near-IR camera consisting of sixteen 2048 $\times$
2048 Raytheon VIRGO HgCdTe arrays distributed over the focal plane of
VISTA. The paw-print of the 16 detectors provides a non-contiguous
instantaneous field of view of 0.6 deg$^2$. Half of detector 16 is affected by a
time-varying quantum efficiency which makes flat fielding that
detector extremely challenging (see \citealt{Jarvis13} for further
details). The NB980 and NB990 filters (Orr, Wallace and Dalton 2008) share a slot on VIRCAM --- each
covering eight detectors, and so to obtain a full pawprint
from either of these filters the survey area has to be observed twice,
with the instrument rotated 180 degrees. Consequently, the two VISTA
NB images suffer from mismatched seeing in the top and bottom halves
of the images, and so we smooth the data to
the PSF of the worst-seeing parts of the VISTA pawprint (1.42 arcseconds). 

\begin{figure}
\begin{center}
\resizebox{0.45\textwidth}{!}{\includegraphics{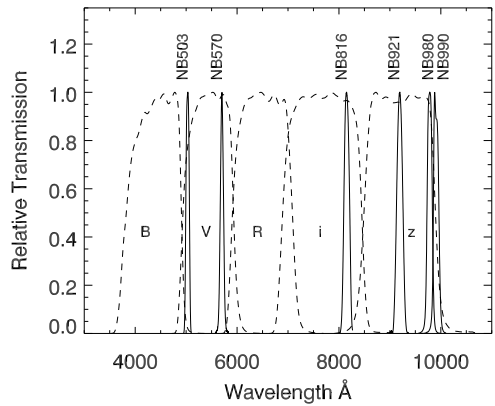}}
\end{center}
\caption[]{Wavelength coverage and transmission functions for all
  selection filters. Four narrow-band filters on Subaru, plus two on
  VISTA (bold black lines) together with broad-band filters at corresponding
  wavelengths also from the Subaru Telescope (dashed lines) are shown.}
\label{fig:filters}
\end{figure}

 In order to select a reliable sample with homogeneously
     determined photometric redshifts, we use only the
     region of sky covered by both SXDS and UDS imaging so as to
     provide 11 bands of photometry for every detection.
The usable overlap of these deep broad-band surveys and the Subaru and
VISTA narrow-band imaging
data amounts to 0.63 sq degrees and 0.38 sq degrees respectively. The
6 narrow-band filters used for selection
are shown together with Suprime-Cam broad-band filters across the corresponding
wavelength ranges in Figure \ref{fig:filters}. 

\section{Sample Construction}
\label{sample}
We use the Source extraction software \textsc{SExtractor}
\citep{Bertin96} to detect objects in each of the narrow-band (NB)
images. \textsc{SExtractor} works by constructing a background map and
then looking for groups of connected pixels above a threshold
calculated from the root mean square (RMS) noise in the image. We require a group of 10
connected pixels at a significance of 1.5 times the background RMS
to confirm a detection. We then 
estimate the narrow-band image depths by
placing 10 000 randomly positioned 2-arcsecond apertures on each
image and iteratively fitting a Gaussian to the sigma-clipped
histogram of resultant fluxes. The catalogue is then cut to demand an
NB-detection above 5$\sigma$ relative
to our estimate of the sky noise in order
for a source to be
selected as a possible line emitter. 

We make a very conservative area cut, dicarding first the official SXDS
masked areas, and then visually inspecting the narrow-band images (with
object detections overlaid) to discard additional areas of noisy
coverage towards the edges of the CCD and then
artifacts, areas around bright stars and CCD bleeds in the broad-band
images. The areas removed from the
study as a result of the visual inspection
varied slightly from filter to filter. 

\subsection{Subaru Selection}
For objects selected in one of the four Subaru NB
filters, the sample was culled to
the 5$\sigma$ limiting magnitude of the shallowest SXDS field. We next
consider the interpolated broad-band colour at the wavelength of the
selection filter as a measure of the continuum level for each object. We
calculate this continuum (IntNB$_{\rm{cont}}$) by interpolating between the two broad-bands adjacent
to the NB assuming that continuum flux follows a
power law: 

\begin{equation}
\rm{IntNB}_{\rm \, cont} = \rm{BB1} - \rm{(BB1-BB2)} \, \frac{\log(\lambda_{\rm
    NB}/\lambda _{\rm BB1} )}{\log(\lambda_{\rm BB2}/\lambda _{\rm BB1} )}.
\label{eq:intnb}
\end{equation}

 As filter NB921 lies
centrally in the \emph{z}-band, we calculate colour in this filter
relative to the \emph{z}-band only. Broad-band magnitudes are obtained by running
\textsc{SExtractor} in dual image mode, and colours are measured in 2 arcsecond
apertures. Objects are expected to scatter around zero in (IntNB$_{\rm{cont}}-$NB),
with emission-line sources showing a positive colour.  In practice,
the median colour of NB-selected sources is slightly offset from zero, due to
the mismatch of interpolated broad-band magnitude and the true continuum magnitude
at the wavelength of the NB. Furthermore, as we select objects across a broad range of NB
magnitudes, we are essentially sampling different redshifts (in addition
to different parts of the galaxy population) and
so with changing NB-magnitude, we introduce a
systematic error that gives rise to a gradient in the median colour of
objects (typically of order $\sim 0.015$ mag). We remove the colour gradient by calculating the
sigma-clipped median colour of sources as a linear function of NB mag, and
subtracting this so that the resultant locus of colours lies along zero. 

\begin{figure*}
\resizebox{0.49\textwidth}{7cm}{\includegraphics{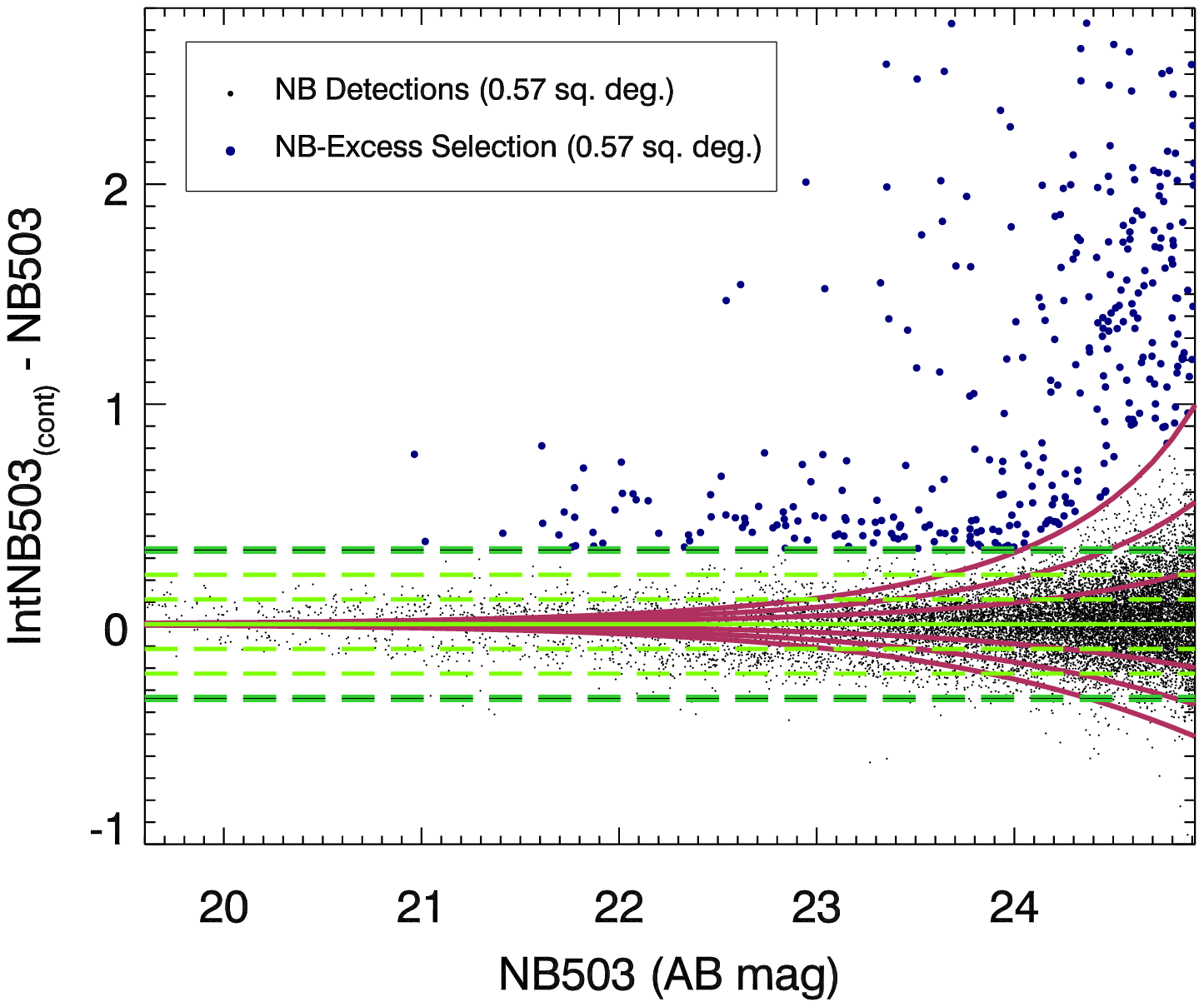}}
\resizebox{0.49\textwidth}{7cm}{\includegraphics{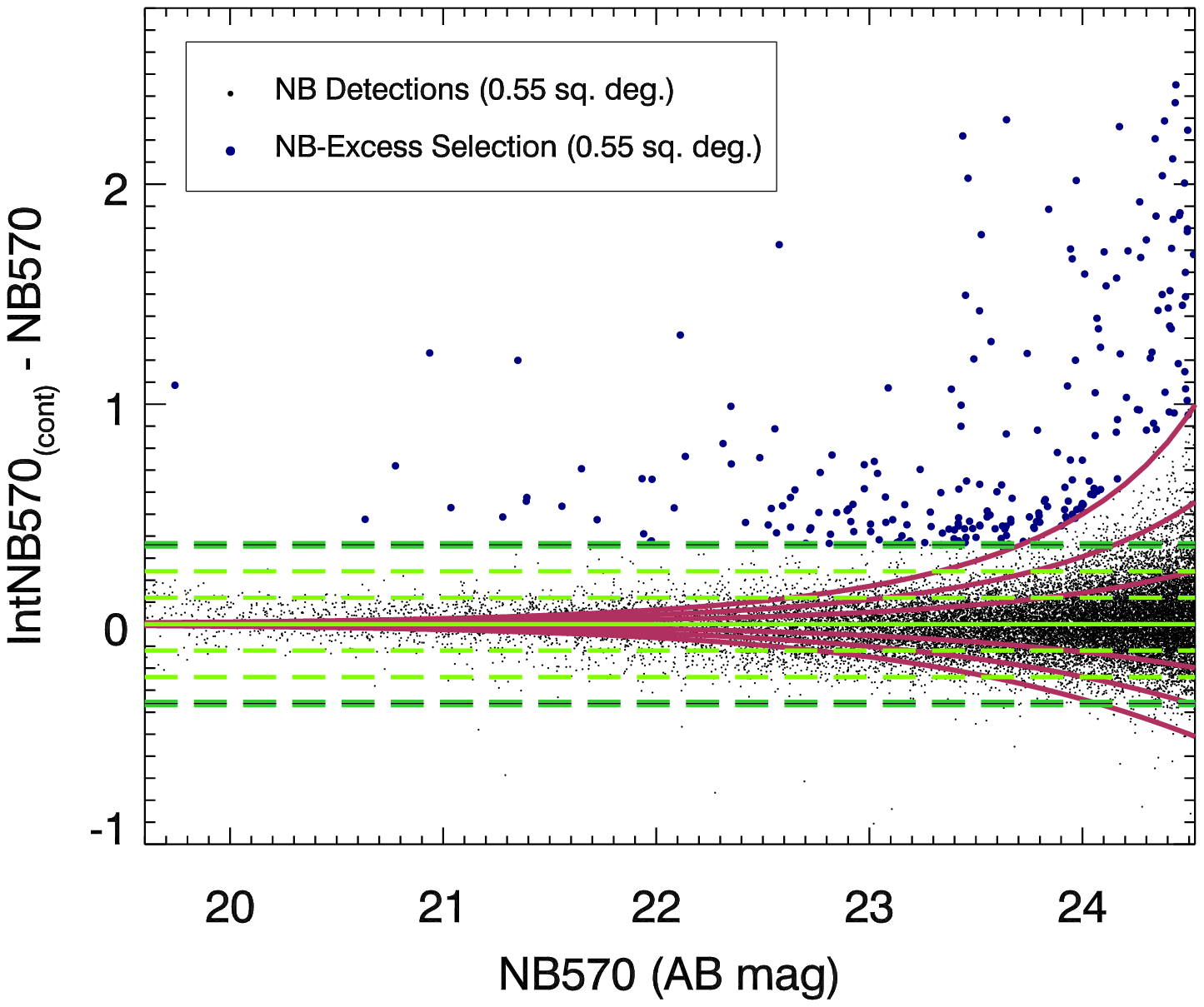}}
\resizebox{0.49\textwidth}{7cm}{\includegraphics{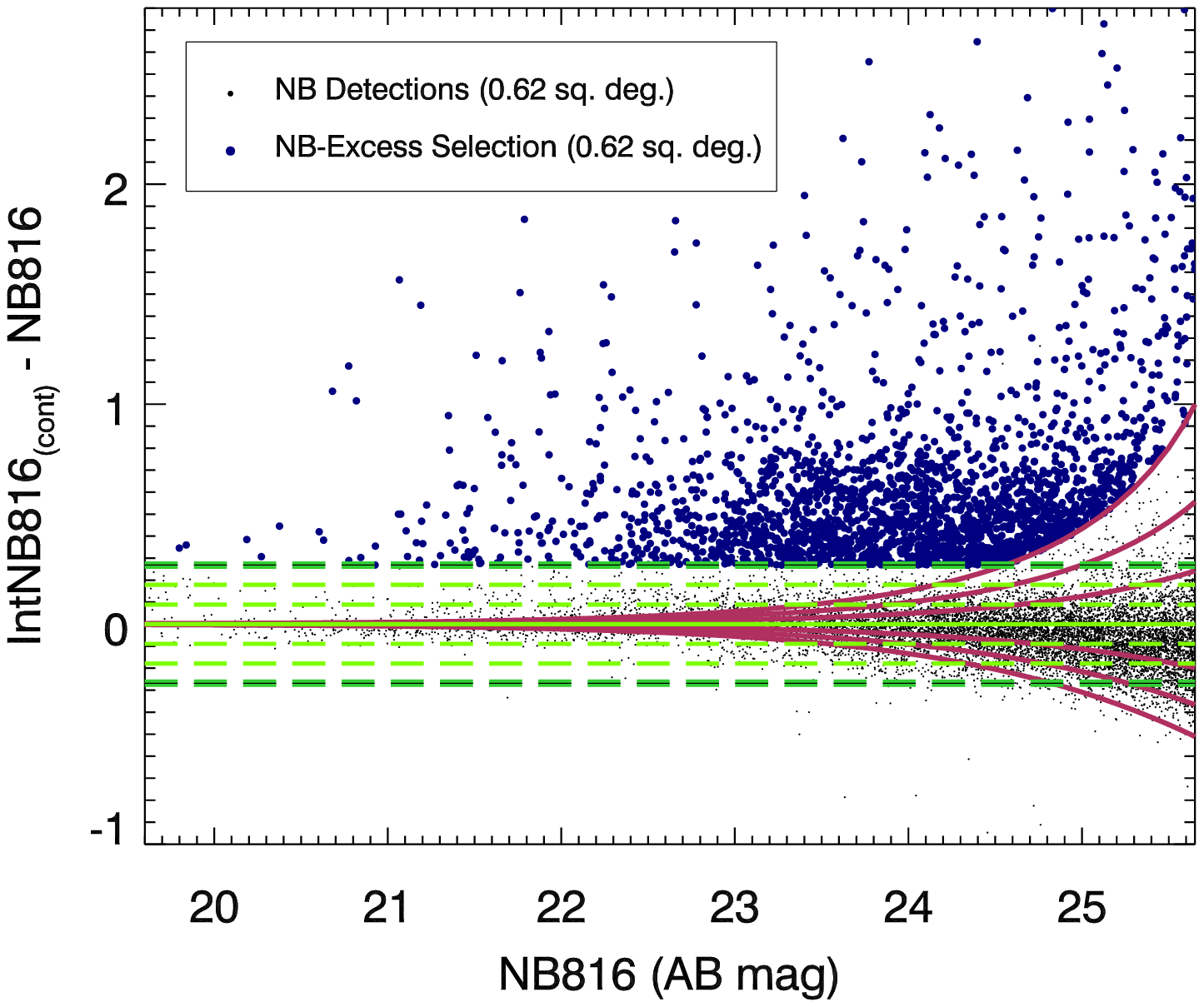}}
\resizebox{0.49\textwidth}{7cm}{\includegraphics{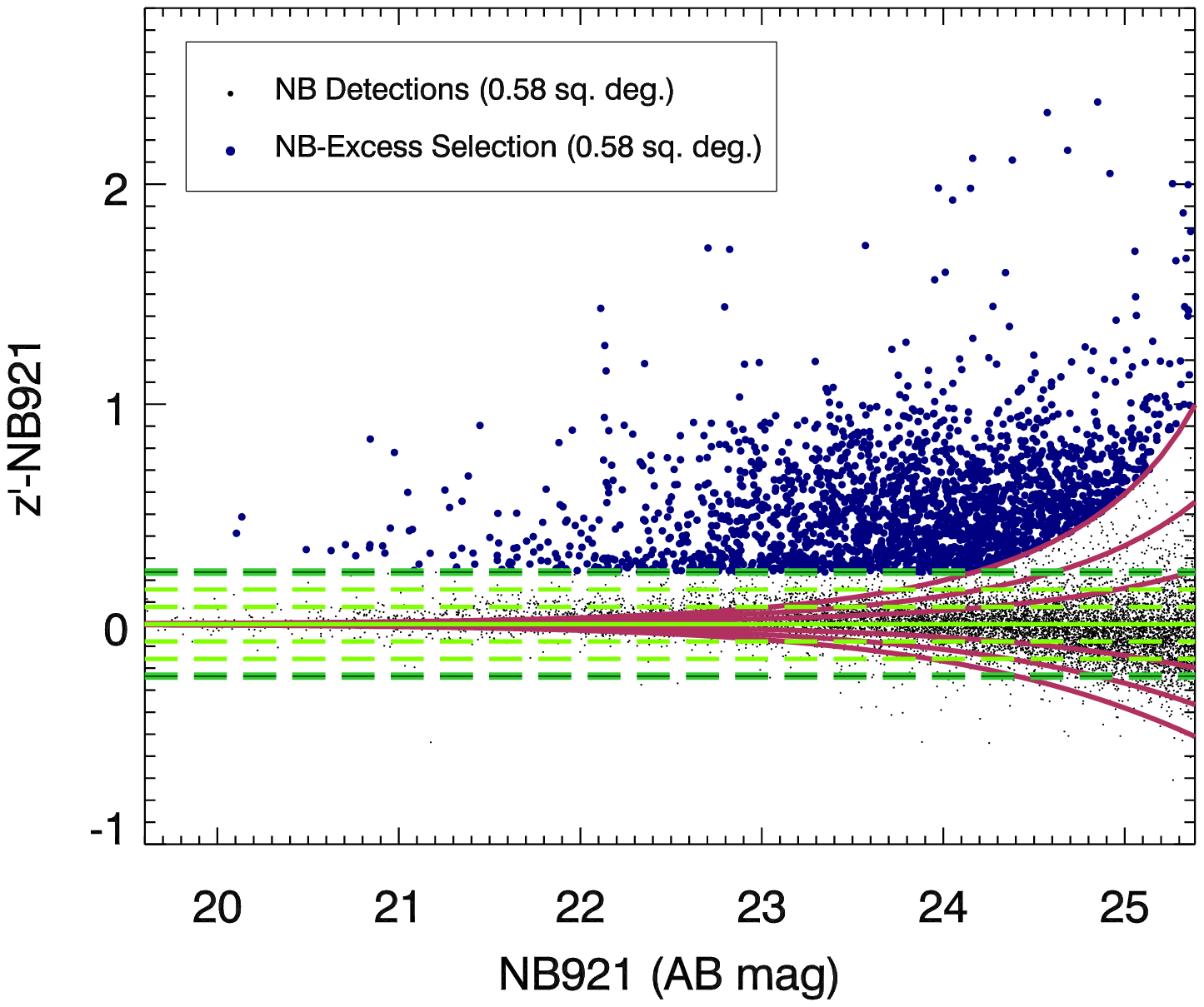}}
\resizebox{0.49\textwidth}{7cm}{\includegraphics{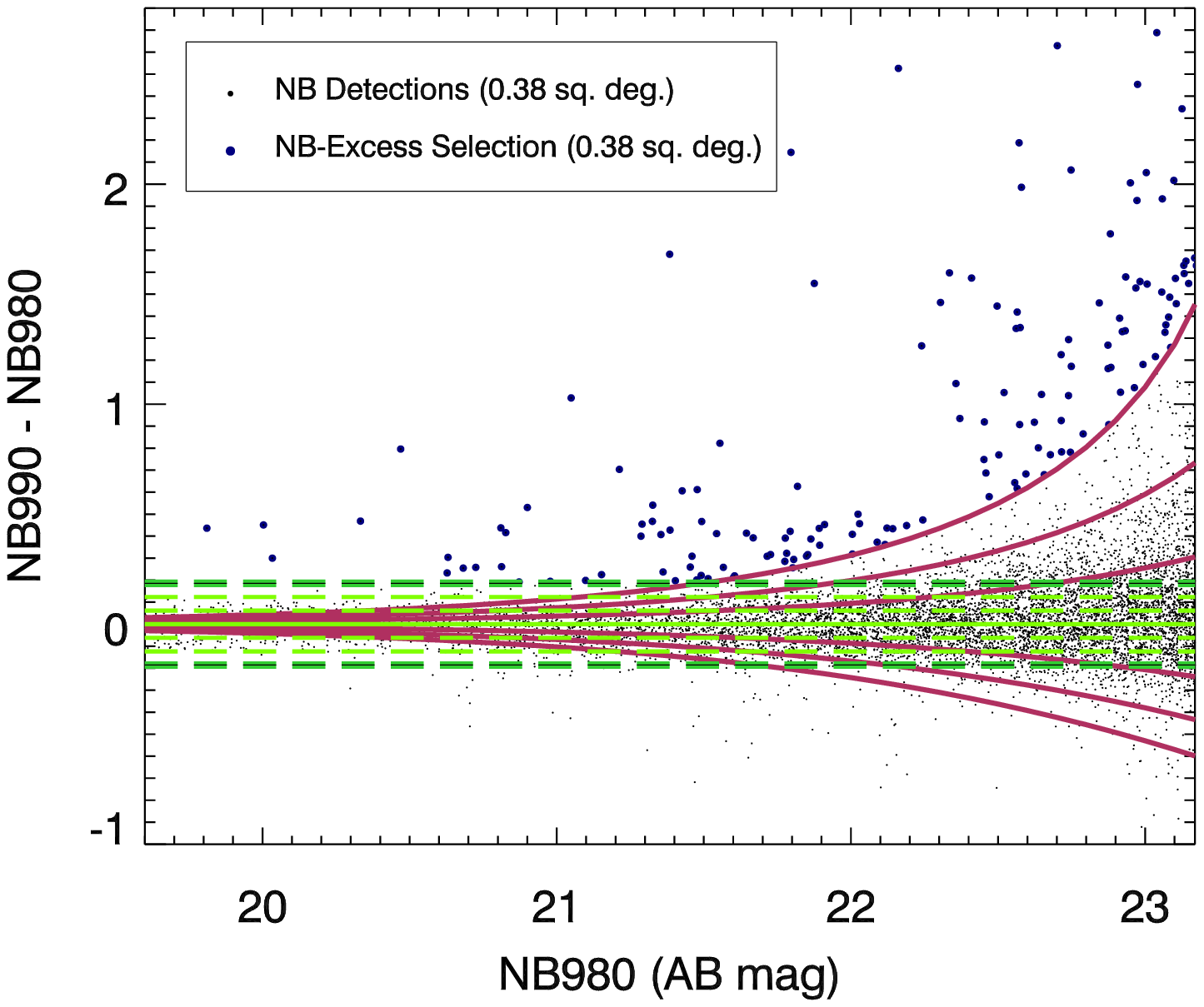}}
\resizebox{0.49\textwidth}{7cm}{\includegraphics{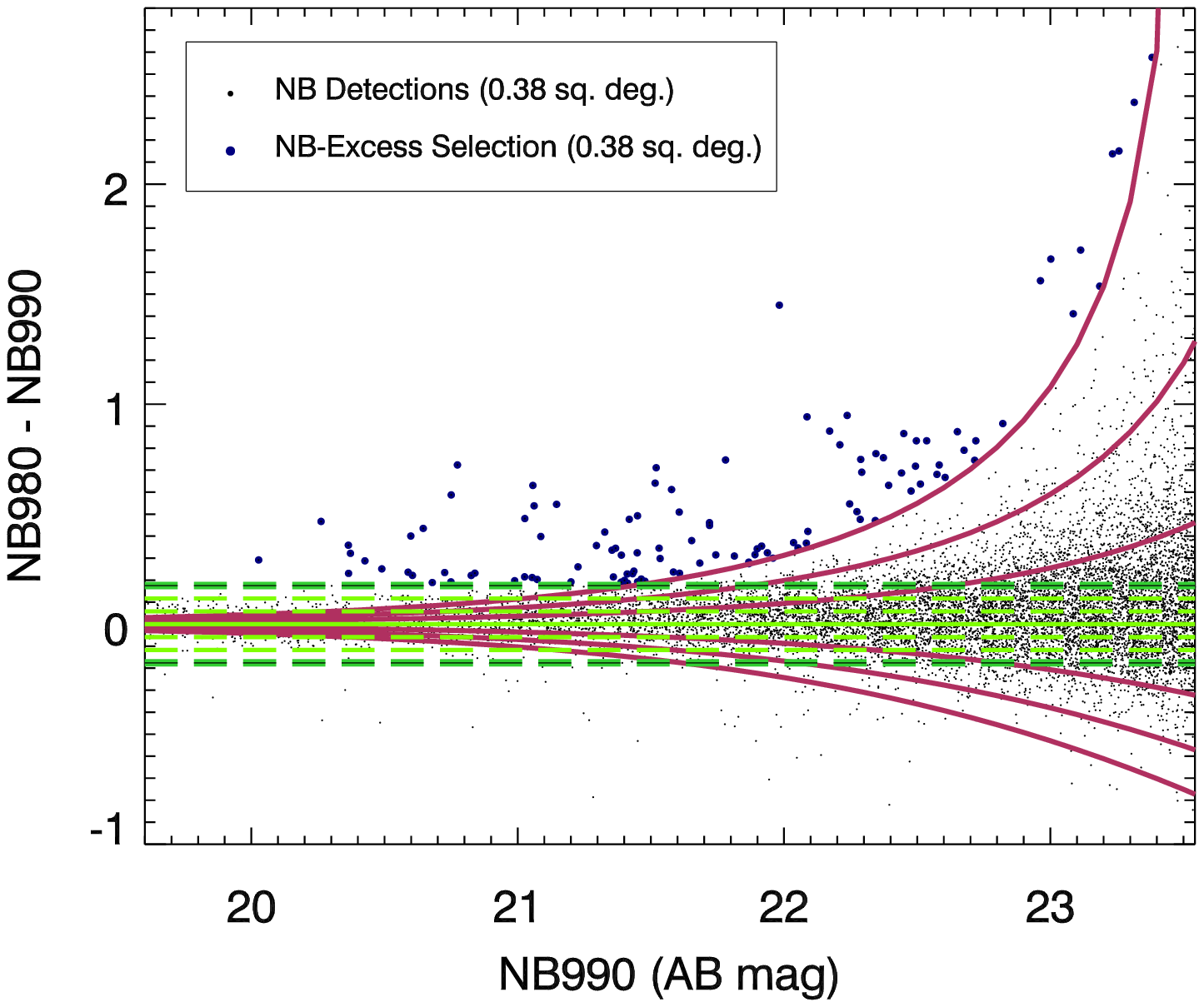}}
\caption[]{The colour--magnitude diagrams
  for the four Subaru narrow-band filters: NB503, NB570, NB816 and
  NB921, and the two VISTA narrow-band filters: NB980 and NB990. Detections are restricted to the 5$\sigma$ limiting NB-magnitude in the
  shallowest SXDS field. The solid green line depicts the median colour of
  sources (corrected for gradient and offset from zero --- see text
  for details), dashed green lines show the 1, 2, and 3$\sigma$ limits on
  intrinsic colour scatter. Solid red lines depict the 1, 2,
  and 3$\sigma$ photometric errors respectively for the shallowest
  region of the five SXDS fields. Solid blue circles represent the
  entire sample of NB-excess-selected objects. }
\label{fig:trumpets}
\end{figure*}

Potential emission-line objects must display an NB-excess of at
    least 3$\sigma$
relative to the sigma clipped median colour {\emph{and}} 3$\sigma$
relative to 
the scatter from photometric uncertainty (in the relevant SXDS field) to be selected as an
emitter. Photometric scatter is determined from the image depth
estimates described above. Figure
{\ref{fig:trumpets}} shows colour-magnitude diagrams for each of the
NB selections, with green lines showing the lines of 1, 2 and 3
$\sigma$ colour excess, and the red lines depicting the 1, 2 and 3
$\sigma$ photometric error trumpets.

We assume our observations are background limited, as even at the bright
magnitudes where photon shot noise comes into play, colour scatter in
(IntNB$_{\rm{cont}}$-NB) is determined empirically and so the
requirement of a 3$\sigma$ excess takes this
scatter into account. 

% TRUMPET DIAGRAMS

\subsection{VISTA Selection}
For selection in one of the two VISTA filters (seen alongside the
Subaru filters in Figure \ref{fig:filters}), detections are made
and NB-excesses evaluated following broadly the same method developed with
the Subaru data. The only variation in the selection process is the
manner in which we evaluate the continuum contribution to the NB
measurement. The two VISTA NBs are close in wavelength, and so they
sample objects at (effectively) the same point in their emission
spectra; measured fluxes should therefore only differ significantly when an
emission line is present in one of the two filters. This allows us to
detect objects in one NB filter, and use photometry from the other as a measure of
continuum flux for each object.

The double-NB selection
technique circumvents the error introduced through inaccurate
determination of interpolated broad-band colour (the main source of scatter introduced into
the NB-broad-band colour) and so we anticipate the locus of VISTA-selected
objects to be much tighter, allowing for a much less conservative
3$\sigma$ cut in colour scatter, i.e. a smaller colour-excess is now
robust at the 3$\sigma$ level. Colour--magnitude diagrams
for these two NBs are also shown in Figure {\ref{fig:trumpets}} demonstrating
the anticipated narrow locus, however the far shallower depth of the
VISTA imaging results in a shallower sampling of the star-forming
population.

%BzK diagrams.
\begin{figure*}
\begin{center}
\resizebox{0.49\textwidth}{7cm}{\includegraphics{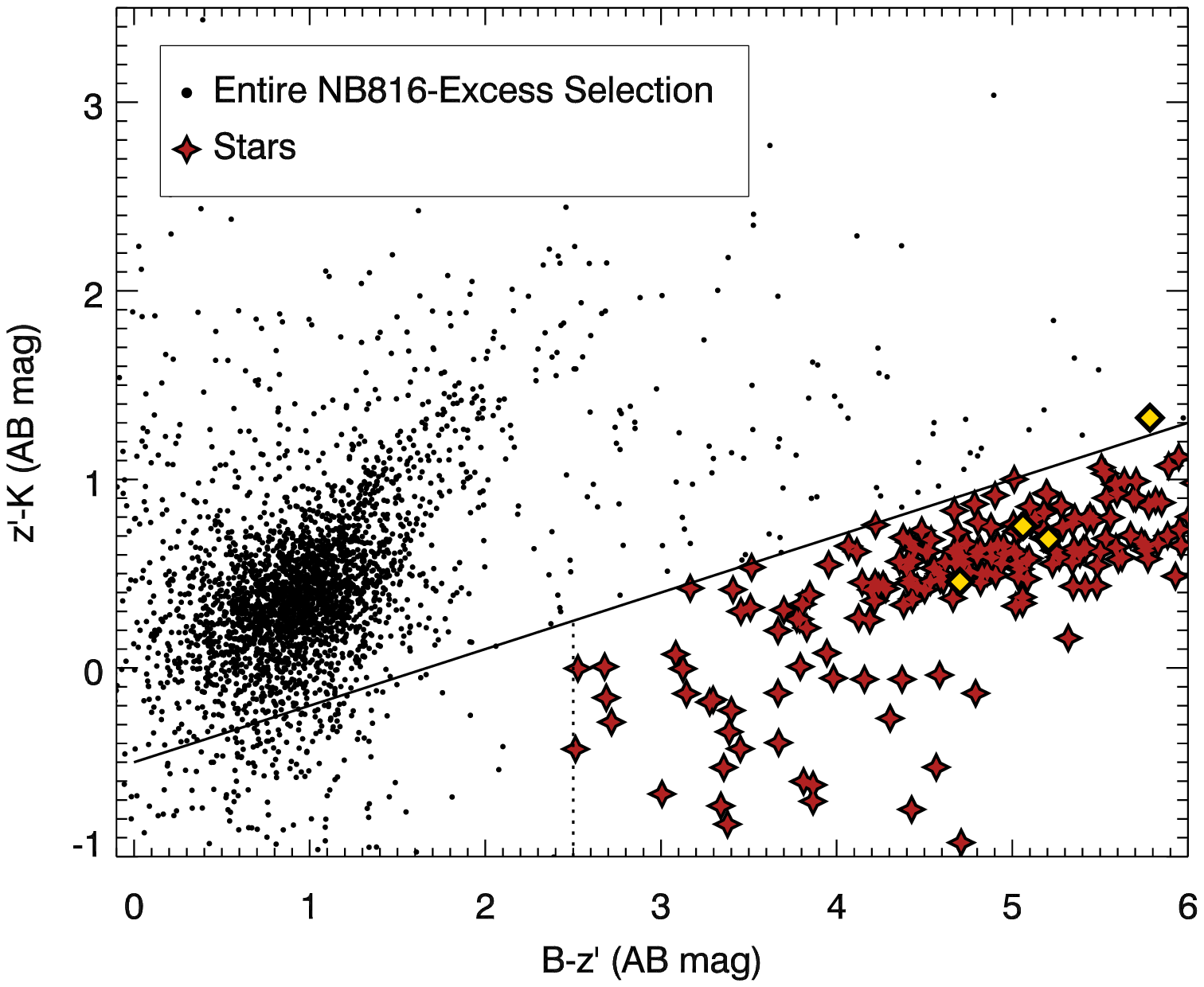}}
\resizebox{0.49\textwidth}{7cm}{\includegraphics{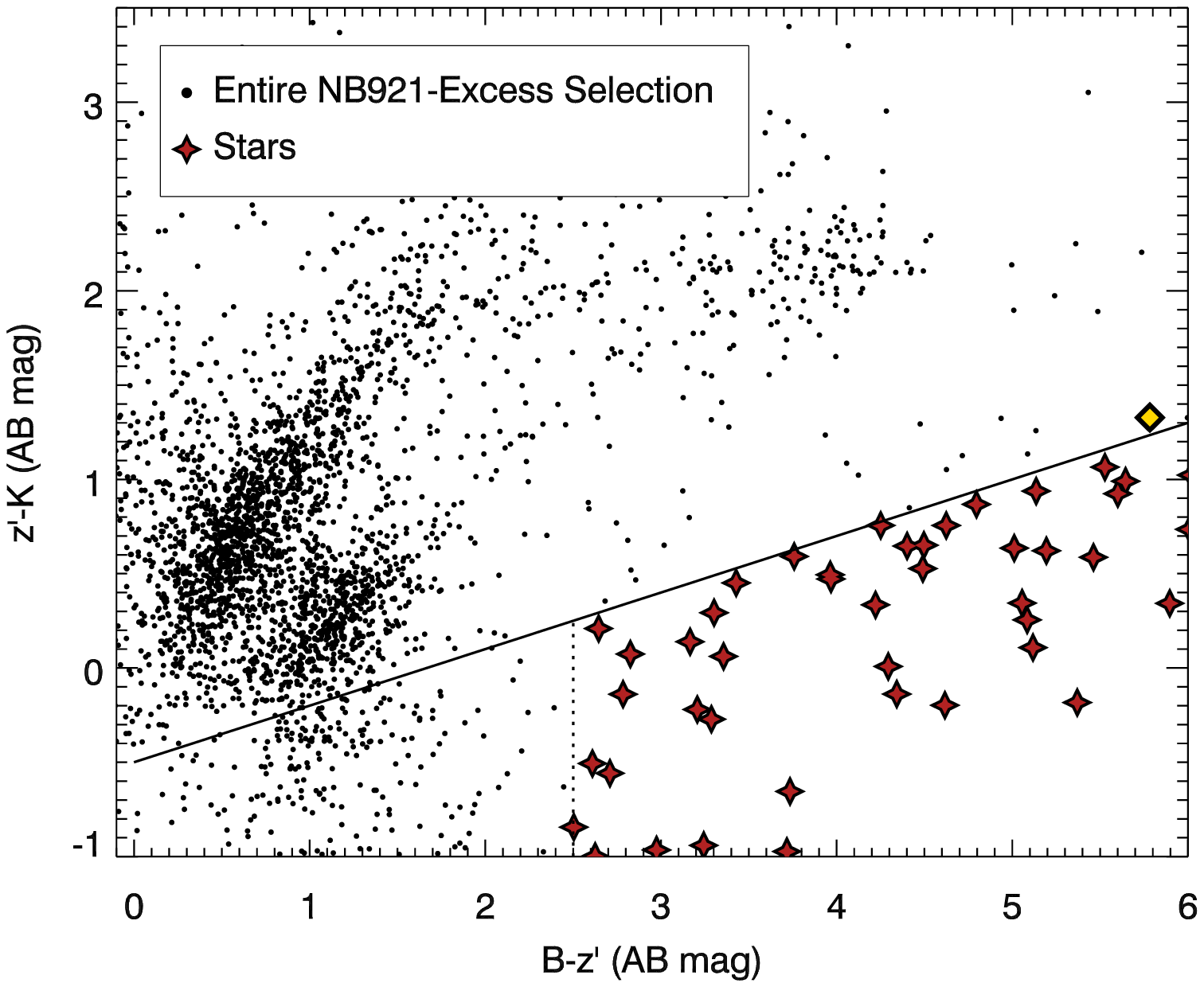}}
\caption[]{\emph{BzK} colour-colour diagrams for the two reddest filters in our Subaru
  NB-Excess selection; NB816 shown on the left and NB921 on the right. All objects falling below the solid black line are classed as
  stars according to \cite{Daddi04}. The subset of these objects that
  we identify as stars (and therefore
  eliminate from our analysis) are highlighted by red stars, yellow
  diamonds represent confirmed stars in the UDS. See main text
  for details.}
\label{fig:BzK}
\end{center}
\end{figure*}

\section{Redshifts}
\label{Redshifts}

Our narrow-band selection, spanning six NB filters from 5029\,\AA\ to 9909\,\AA\, is
sensitive to all the major diagnostic emission lines in the region
from [O{\sc~ii}] to H$\alpha$ + [N{\sc~ii}], and samples
redshift slices from z=0.14 ([O{\sc~iii}] in the NB570 filter) to
z=1.6 where [O{\sc~ii}] falls in to the VISTA filters. With more than 7000
objects meeting the selection criteria, it is essential that we are
able to reliably discriminate between the different lines falling in
each filter while employing a technique suitable for dealing with such
a volume of data. The combination of the deep SXDS-UDS imaging, and complementary \emph{u} band and
IRAC data, provides 11 broad bands of photometry available for every
detection, allowing reliable determinations of objects' spectral
energy distributions (SEDs) and
consequently accurate photometric redshifts.
We have obtained spectra for a small subset of objects (discussed in
Section \ref{spec}) and use the
redshifts derived as a training set for the photometric redshift code
``EAzY'' (Brammer, Van Dokkum and Coppi 2008).

\subsection{Photometry}
\label{sect:phot}
Photometry for the full selection of NB-excess selected objects is
computed in all 11 broad bands using the \textsc{iraf} task \texttt{phot} in 2 arcsecond
apertures. Astrometric corrections between Suprime-Cam data and the
other images were performed using the method of
\cite{Simpson12}. All Suprime-Cam imaging was reduced to provide a uniform PSF of
    $0.82$ arcsec, and the \emph{J}, \emph{H}  and
\emph{K} band images from WFCAM are consistent with this to within
$0.05$ arcsec, and so no corrections are applied to the measured photometry. \emph{B} and \emph{K} band images were smoothed using a Gaussian kernel to
recover the same PSF as the \emph{u} and IRAC images respectively, and fluxes from the \emph{u} band
image and the two IRAC images were then scaled according to the ratio
of fluxes in the unsmoothed and smoothed images. All broad-band
photometry is corrected for Galactic extinction of $A_V = 0.070$ magnitudes \citep{Schlegel98} using
the  \cite*{CardelliClaytonMathis} reddening law.

\subsection{Star--Galaxy Separation}
\label{stargal}
Our narrow-band-excess technique is inherently
susceptible to the spurious selection of any object with a spectrum
not well described by a power law. This leads to contamination of the
sample by late-type stars which are known to exhibit very red, complex
SEDs due to the many absorption features present
in their spectra. We employ the two-colour \emph{BzK} technique of
\cite{Daddi04} to identify and eliminate these stars from the red
filters susceptible to this contamination. For the Subaru NB filters
this applies to NB816 and NB921. Daddi et al., (2004)
derive Equation \ref{eq:Daddi}, and class all objects below this line as
stars. 

\begin{equation}
{(\emph{z} -\emph{K}) < 0.3\,(\emph{B} -\emph{z}) -0.5}
\label{eq:Daddi}
\end{equation}

As we are contaminated only by late-type, red, stars, we
  employ the additional criterion that objects must have \emph{B}-\emph{z} \textgreater 2.5
 to be classed as stars and eliminated from the
  selection. 

The \emph{BzK} colours of our sample are shown in Figure \ref{fig:BzK}. The
population of blue objects falling below the \cite{Daddi04} \emph{BzK} solid
black line arises from scatter as a consequence of very faint \emph{K} magnitudes
   ($>$ 26). We eliminate 256 stars in  total, amounting
   to 2 per cent of the NB-excess sample. 

\begin{table}
\caption{Detections per NB filter. Excess objects are defined as objects
  presenting an NB-excess above 3$\sigma$ in intrinsic colour scatter
  and photometric error. Galaxies are those objects that remain
  following the elimination of stars via our star--galaxy
  separation technique (see Section \ref{stargal}).}
\label{tab:numbers}
\begin{center}
\begin{tabular}{cccc} \hline
Filter & Detections & Excess Objects & Galaxies\\
\hline
NB503 & 43781 & 4368 & 4368\\
NB570 & 37846 & 1749 & 1749 \\
NB816 & 97382 & 3816 & 3595 \\
NB921 & 109126 & 5395 & 5360 \\
NB980 & 16193 & 119 & 119 \\
NB990 & 17755 & 191& 191 \\
\hline
\end{tabular}
\end{center}
\end{table} 

Despite the red wavelength of the two VISTA filters, only NB990 is in
danger of the spurious selection of stars due to the point in a late
type star spectrum it samples. Running the NB990 selection through our
\emph{BzK} criteria however eliminates nothing from the small sample. Table
\ref{tab:numbers} summarizes the number of detections in each filter
plus numbers remaining in the sample after quantifying NB-excess and
removing stars from the sample.

\subsection{Spectroscopic Subset Analysis}
\label{spec}
With a preliminary selection of NB-excess objects in the central field
of the SXDS, we compiled a target list for follow-up
spectroscopy. Observations were carried out with the Inamori Magellan
Areal Camera Spectrograph (IMACS)
instrument on the Walter
Baade Telescope on the dates of 2011 Jan 03-04, using 0.8 arcsecond
slits and the 200 lines per mm grism (covering the wavelength range
$\Delta \lambda = 0.39 - 1.00 \mu$m, giving
$\Delta\lambda \sim 8$ \AA\ pix$^{-1}$ at FWHM). The total exposure time was
$335$ minutes, and seeing varied from $0.6 - 0.9$ arcseconds.

The spectra were reduced using the COSMOS reduction pipeline before
extracting redshift information using the \emph{runz}
redshifting code \citep{runz}. We determine reliable
redshifts for 129 of the sample, and assign 108 of these the highest
quality flag. The remainder of the subset is made up of low
signal to noise detections (where emission-line flux does not in fact meet our final
criteria), spectra displaying a single unidentifiable line, spectra
affected by cross-talk from other slits, and a
small number of stars targeted to confirm our star--galaxy separation
technique. 

The modest but reliable subset of spectroscopic
redshifts confirms that we are indeed selecting emission-line
galaxies at the redshifts anticipated, and serves as a reference to
determine the success of our photometric redshift technique applied to
the rest of the sample.

\subsection{Photometric Redshift Analysis}
\label{PRA}
The default parameters of EAzY provide photometric redshifts in good
agreement with the distribution expected for emission-line objects
selected in these filters; however, to optimize the SED-fitting, we
modify this default parameter set so as to minimize
$\sigma_{\rm{NMAD}}$ for the spectroscopic subset (Equation
{\ref{eq:sigNMAD}}; \citealt{BrammerVanDokkumCoppi}), where 

\begin{equation}
{\sigma_{\rm{NMAD}} = 1.48 \times {\rm{median}} \left(\left|\frac{\Delta z-{\rm{median}}(\Delta z)}{1 +  z_{\rm{spec}}}\right|\right).}
\label{eq:sigNMAD}
\end{equation}

\begin{figure}
\begin{center}
\resizebox{0.95\hsize}{7cm}{\includegraphics{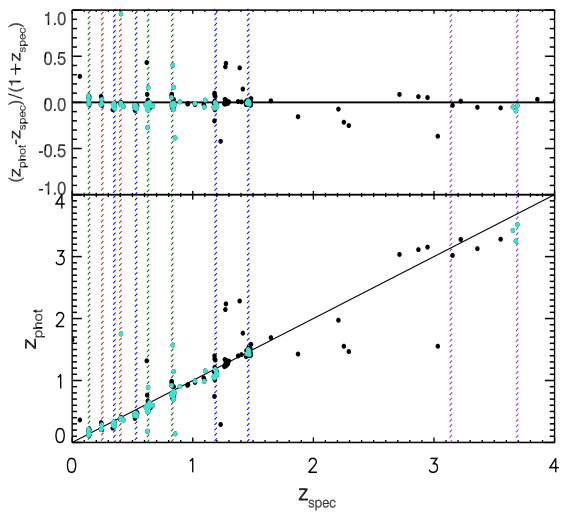}}
\caption[]{Comparison of spectroscopic and photometric redshifts for
  the subset of objects with spectroscopic follow-up, and spectroscopic
  redshifts of the highest
  quality flag. The bottom panel shows $z_{\rm{phot}}$ against $z_{\rm{spec}}$ for the optimized set of EAzY
  parameters. The top panel shows the distribution of $\Delta$z
  around zero. Cyan points represent our Magellan spectroscopic data,
  and black points represent all other available spectra in the UDS.}
\label{fig:zphotzspec}
\end{center}
\end{figure}

The number of catastrophic outliers ($>$ 5$\sigma$$_{\rm{NMAD}}$) for
definite spectroscopic redshifts (of the highest quality flag)
equates to 5 per cent of the spectroscopic subset, with a value of
$\sigma$$_{\rm{NMAD}}= 0.026$. Including objects
with less certain spectroscopic redshifts raises this to 9
per cent. While photometric redshifts are commonplace for large survey data sets, particularly
for \emph{K}-selected (i.e. stellar-mass-limited) samples, photometry is
frequently restricted to broad-bands only. Here we possess additional
information that all objects remaining after the
selection process must display a narrow-band excess at the wavelength
of the filter. 

While na\"{\i}vely, we might interpret the presence of a narrow-band excess
to mean that all sources must possess a major emission line redshifted to the
wavelength of the selection filter, in practice the problem proves
to be more complex. We design a pure emission-line template to
investigate  the effect of including narrow-band data in addition to
broad-band photometry, before fine
tuning the setup parameters. Our emission line template 
consists of the major emission lines Ly$\alpha$, [O{\sc~ii}],
H$\beta$, [O{\sc~iii}], H$\alpha$ and [N{\sc~ii}] with negligible continuum
flux. However, including this template now means that the complexities of the emission-line ratios come
into play, along with the nature of the narrow-band excess. While the
objects do indeed lie in distinct redshift slices, narrow-band excess can
arise from a vast number of emission lines. We find that the inclusion of narrow-band data, along
with an emission-line template, biases redshift determination towards
the redshift slices associated with the emission lines in our
template at the \emph{expense} of a good fit to the broad-band
photometry, thus adding redshift interlopers to our selection. 
The 11 bands of photometry are deep, and as EAzY is well equipped to
determine redshifts based on this information alone we discard the
narrow-band plus emission-line template setup at this stage. 

 We find the best agreement between  $z_{\rm{phot}}$ and
 $z_{\rm{spec}}$ for the default template set, using broad-band data
 only (see Figure \ref{fig:zphotzspec}). This way redshifts are
 assigned independently of their detection method, allowing
 the correct identification of spectral breaks and additional better
fitting emission lines than those included in the emission-line
template. For the comparison in Figure \ref{fig:zphotzspec} we also incorporate matches to
    the full set of spectra available in the UDS field. The larger
    SXDS--UDS spectroscopic sample (\citealt{Simpson12}, and papers cited
    therein) includes many spectra that are the result of
    targeted follow-up of X-ray or radio sources, and is therefore heavily biased towards Active
Galactic Nuclei (AGN). 
  Objects with spectra from the larger sample (depicted by black dots in Figure
  \ref{fig:zphotzspec}) often appear at redshifts other than those
  typical for our selection, as they sample many
    objects with emission lines usually weak in star--forming galaxies.

\subsection{{AGN Contamination}}
 We identify 8 (of the 129) objects as AGN according to the line-ratio diagrams of
\citet{Lamareille10} or based on the presence of [Ne{\sc~iii}] emission. As we can use
    only our own spectroscopic follow-up from the Magellan telescopes
    to quantify AGN contamination in an unbiased manner, we infer a contamination fraction
    of 6$^{+8.9}_{-3.9}$ percent. This correction is small and highly
    uncertain, and because AGN activity is undoubtedly associated with some star
    formation, removing these sources then leaves us
    susceptible to a possible {\emph{over}}--correction and consequently
    an underestimate of star formation activity. We
therefore choose not to correct our results for AGN contamination.

\subsection{Resultant Photometric Redshift Distribution}

\begin{figure}
\begin{center}
\resizebox{0.48\textwidth}{!}{\includegraphics{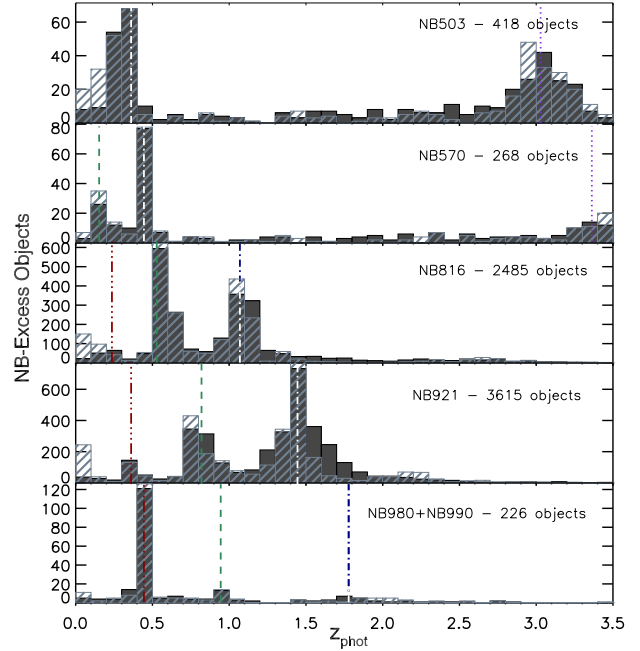}} 
\caption[]{Photometric redshift distributions for the 4 Subaru
  narrow-band filters and the combined distributions from VISTA's NB980
  and NB990 filters. The dark grey filled histogram represents $z_{\rm{m2}}$, the
  redshift marginalized over the posterior redshift probability
  distribution, and the light grey hatched histogram traces $z_{\rm{a}}$, the
  best fit redshift. Panels top to bottom: NB503, NB570, NB816, NB921
  and NB980+NB990 in redshift bins of width $\Delta z$= 0.1. Vertical lines
  indicate the redshifts at which different emission lines would be
  detected in the various filters: triple-dot-dashed lines show
  H$\alpha$ (dark red), dashed lines the [O{\sc~iii}] doublet and
  H$\beta$ (green), and  dot-dashed lines [O{\sc~ii}] in
blue. Filters NB503 and NB570 also show peaks at z$\sim$3 where we detect Ly$\alpha$ (dotted purple line).}
\label{fig:zphots}
\end{center}
\end{figure}

EAzY provides an output probability density distribution $P(z)$ for each
object it evaluates. We use  $z_{\rm{m2}}$, the redshift marginalized over the posterior redshift probability distribution
(also taking into account the \emph{K}-band apparent magnitude), to determine the redshift distribution of our objects
and discriminate between the emission lines giving rise to narrow-band-excess. The
resultant photometric redshift distribution can be seen in Figure
{\ref{fig:zphots}}. The advantage of the posterior redshift
probability is two-fold: we combine the best estimate of redshift
(using all SED templates) with \emph{a priori} constraints on the likelihood
of an object being at a particular redshift (given our knowledge of its
colour and \emph{K} magnitude prior), while taking into account the
probability that the object lies at an alternative, less well-fit
redshift. $z_{\rm{m2}}$ effectively provides a weighted mean, however the
resultant distribution of redshifts and that for the best-fitting
redshift ($z_{\rm{a}}$) are in
fact very similar (traced by the two histograms in Figure \ref{fig:zphots}), which demonstrates
that we are not altering the best-fitting redshifts dramatically through
marginalisation and
use of the prior. We note that the small number of catastrophic
outliers discussed in Section \ref{PRA} produce
doubly-peaked $P(z)$ distributions, with $z_{\rm{a}}$ often selecting a redshift of relatively
small probability compared to the probability at the value of
$z_{\rm{m2}}$. Similarly, the small peaks of objects lying at $z_{\rm{a}}$=0 are the
result of objects missing a red band of photometry and producing a small
peak at $z_{\rm{a}}$=0 in their $P(z)$ distributions. This lends further support to the
advantages of $z_{\rm{m2}}$ which in these cases avoids the small $z_{\rm{a}}$=0 spike via
the weighted mean. Figure \ref{fig:zphots} shows that we are selecting
emission-line galaxies at the anticipated redshifts and highlights
the peaks in the distribution associated with  H$\alpha$, [O{\sc~iii}]
and [O{\sc~ii}], in addition to Ly$\alpha$ in the two bluest filters
(\citealt{Ouchi08}; not included in this analysis). 

The lowest panel of Figure \ref{fig:zphots} shows the combined
data sets of NB980 and NB990. As these two filters are very
close in wavelength ($\Delta\,\lambda_c$ $\sim$120\AA) we choose to combine the two data
sets to combat the small number of detections (with a 5$\sigma$ detection cutoff in
the shallowest filter; NB980). The NB980 and NB990 data sets are
evaluated separately, and then combined before assigning objects to a
redshift slice. The combined data set (hereafter NB985) is treated
as a single selection from here on.

\subsection{Redshift Slice Assignment}
\begin{figure}
\begin{center}
\resizebox {0.48\textwidth}{!}{\includegraphics{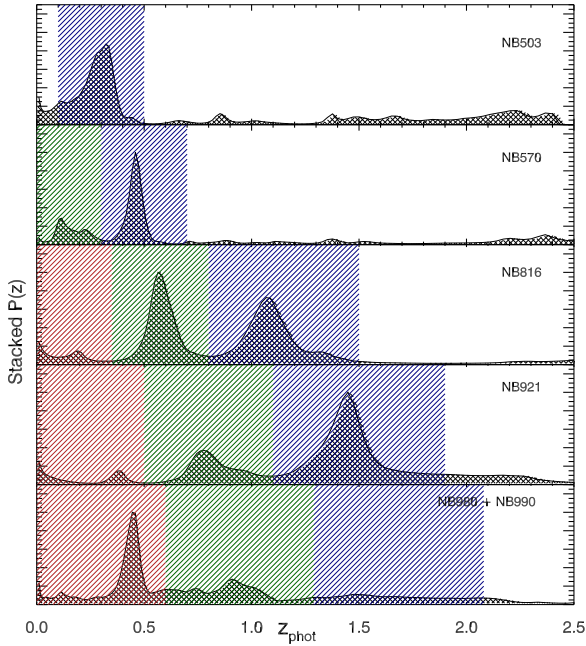}} 
\caption[]{Stacked probability density distributions for each
  NB-selection of objects, shown over the redshift range relevant for
  this study. Red (H$\alpha$), green ([O{\sc~iii}]) and
  blue ([O{\sc~ii}]) shaded regions
  represent the redshift ranges assigned to a slice for a particular
  filter/line combination.}
\label{fig:stack}
\end{center}
\end{figure}

Before assigning objects to a particular redshift slice we consider
the stack of $P(z)$ values for each NB-selection of objects (seen in Figure
{\ref{fig:stack}}). We bin objects into the redshift slices associated
with the peaks in the data, and assign the upper and lower limits of
the bins according to the minima in the stacked $P(z)$ distribution. For the VISTA selection where the minima of the distribution are less
apparent, the widths of the NB921 redshift windows are applied, centred
on the peaks of the VISTA data. Table \ref{tab:slices} summarizes the final numbers of objects in each
redshift slice and the redshift ranges over which each peak was
defined.

\begin{table}
\caption{Number of objects assigned to each redshift slice. The
  redshift range quoted refers to the limits in $z_{\rm{m2}}$ according to
  which objects were binned
  into redshift slices. These limits are taken from the minima in
  Figure \ref{fig:stack} with the exception of the VISTA data for which the
  NB921 limits were transposed to the appropriate redshift. The number of
  objects per slice is the number of real objects assigned to a
  redshift slice from our catalogue, that were used to derive a
  luminosity function for this redshift range.}
\label{tab:slices}
\begin{center}
\begin{tabular}{cccc} \hline
Filter & Redshift & Line & Objects \\
\hline
NB503 & 0.10 $<$ {\bf{0.35}} $<$ 0.50& [O{\sc~ii}] & 142 \\
\hline
NB570 & 0.00 $<$ {\bf{0.14}} $<$ 0.30 & [O{\sc~iii}] & 42 \\
NB570 & 0.30 $<$ {\bf{0.53}} $<$ 0.70 & [O{\sc~ii}] & 96 \\
\hline
NB816 & 0.00 $<$ {\bf{0.25}} $<$ 0.35 & H$\alpha$ & 152 \\
NB816 & 0.35 $<$ {\bf{0.63}} $<$ 0.80 & [O{\sc~iii}] & 985 \\
NB816 & 0.80 $<$ {\bf{1.19}} $<$ 1.50 & [O{\sc~ii}] & 1013 \\
\hline
NB921 & 0.00 $<$ {\bf{0.40}} $<$ 0.50 & H$\alpha$ & 279 \\
NB921 & 0.50 $<$  {\bf{0.83}} $<$ 1.10 & [O{\sc~iii}] & 930 \\
NB921 & 1.10 $<$ {\bf{1.46}} $<$ 1.90 & [O{\sc~ii}] & 2204 \\
\hline
NB985 & 0.00 $<$ {\bf{0.50}} $<$ 0.60 & H$\alpha$ & 153 \\
NB985 & 0.60 $<$ {\bf{0.99}} $<$ 1.29 & [O{\sc~iii}] & 33 \\
NB985 & 1.29 $<$ {\bf{1.64}} $<$ 2.08 & [O{\sc~ii}] & 23 \\
\hline
\end{tabular}
\end{center}
\end{table}

\subsection{Line fluxes}
Narrow-Band selections are by necessity made in an observable
parameter space, here based on line flux. Apparent aperture line
fluxes (hereafter: line fluxes) for objects in
each redshift slice are derived from the initial \textsc{SExtractor} detection
measurements. The entirety of the NB-excess is assumed to arise from
an emission line and so line flux ($\mu$Jy)  is computed as Equation
\ref{eq:lineflux}, and converted to Wm$^{-2}$.

\begin{equation}
 {{\rm{Line\: flux}}= {\rm{flux_{NB}}} -{\rm{flux_{cont}}}}
\label{eq:lineflux}
\end{equation}

\noindent where flux$_{\rm{NB}}$ is the flux in the narrow-band, and flux$_{{\rm{cont}}}$ is
the continuum contribution to that narrow-band flux.

\begin{figure*}
\begin{center}
\resizebox{0.3\textwidth}{5.5cm}{\includegraphics{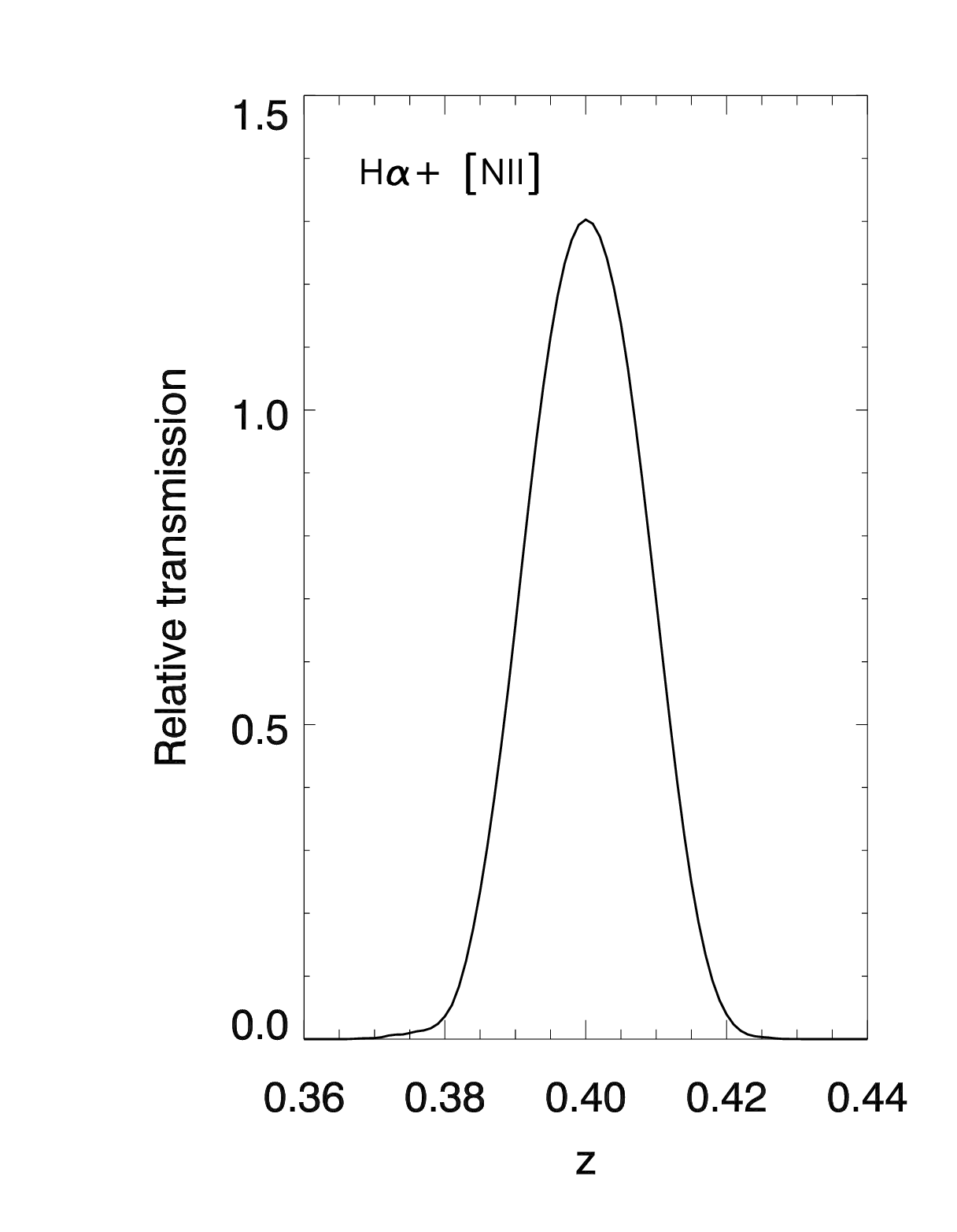}}
\resizebox{0.3\textwidth}{5.5cm}{\includegraphics{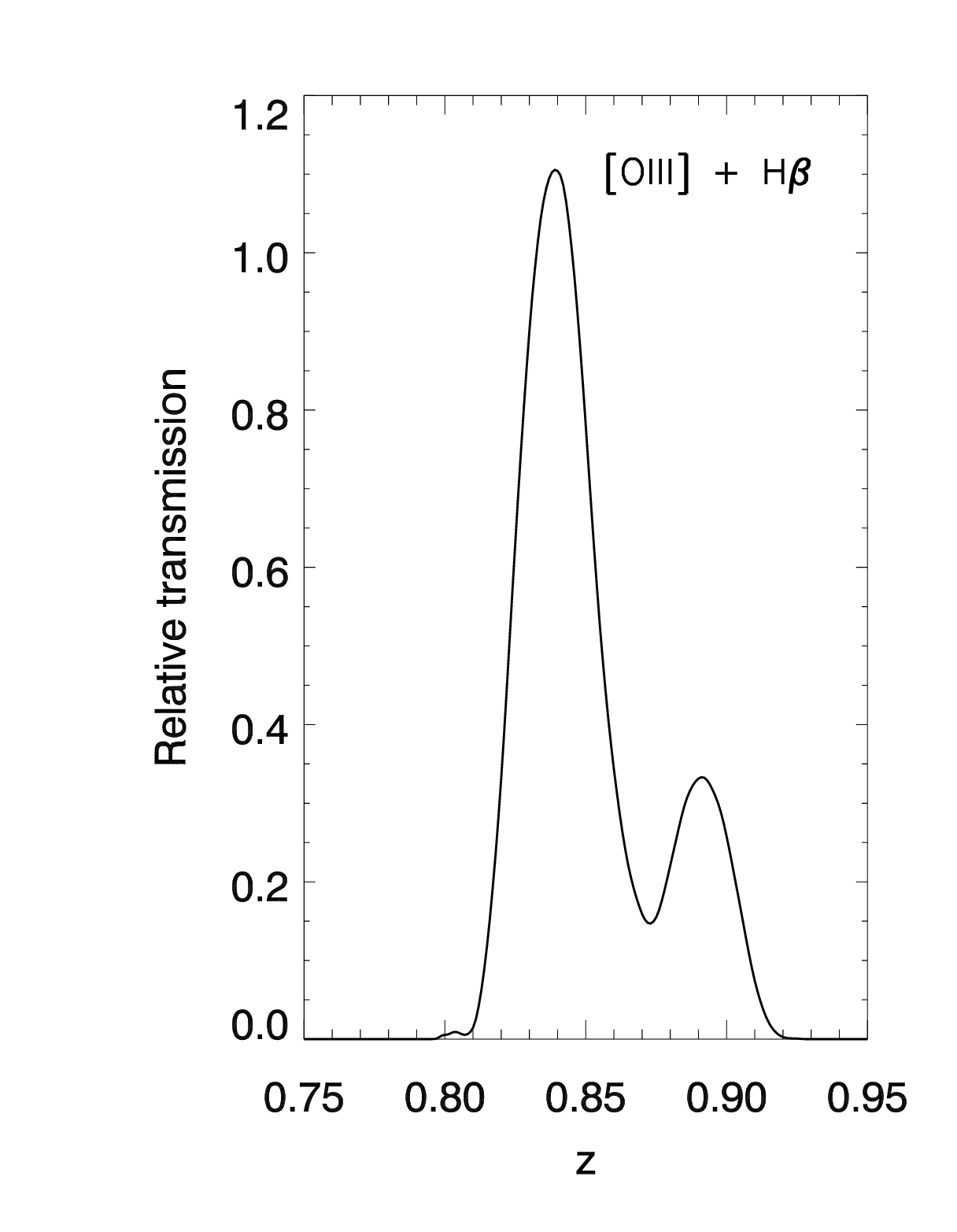}}
\resizebox{0.3\textwidth}{5.5cm}{\includegraphics{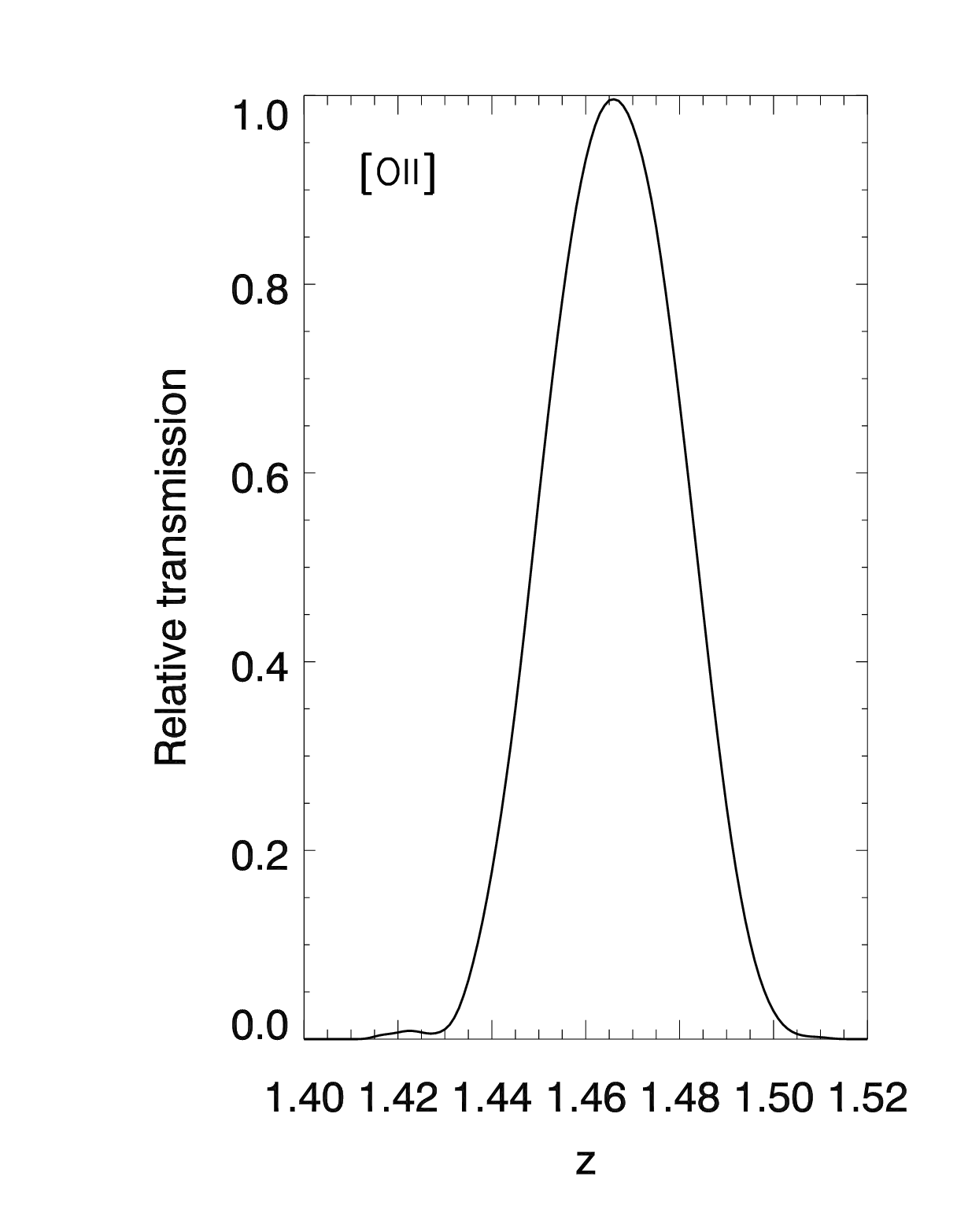}}
\caption[]{Line transmission for NB921. Panels from left to
  right show the effective line transmission with redshift for H$\alpha$, [O{\sc~iii}] and
  [O{\sc~ii}] respectively, as a synthetic spectrum is propagated
  through the NB921 filter.}
\label{fig:filttrans}
\end{center}
\end{figure*}

\section{Maximum Likelihood Luminosity Functions}
\label{ML}
We employ the maximum likelihood method of
\cite {Marshall1983} in order to fit luminosity functions to the
data. We assume the distribution of line luminosities in a given
redshift slice will follow a Schechter function, given in log form by
Equation \ref{eq:Schechter}:

\begin{equation}
 {\phi\,(L)\,{\rm{d log}}L= {\rm{ln10}} \, \phi^{*}
   \left(\frac{L}{L^{*}} \right) ^{\alpha +1} \, e^{-({L}/{L^{*}})}\,{\rm{d log}}L} 
\label{eq:Schechter}
\end{equation}

\noindent where $\phi$$^{*}$, $L^{*}$ and
$\alpha$ are the characteristic number density,
characteristic luminosity,  and the gradient of the
faint-end slope respectively \citep{Schechter76}. We assume negligible
cosmic evolution across the small redshift interval per slice.
Next we consider how the distribution
of true line luminosities maps onto our observed distribution of
line fluxes. 

Splitting the line flux range where we are sensitive to lines into
bins small enough to expect no more than 1 object per bin, we can
write the likelihood of finding an object in bins {F$_i$} and no
objects in bins {F$_j$}, as Equation
\ref{eq:L} for a given luminosity function:

\begin{equation}
{ {\Lambda} = {\prod_{F_i} } \Psi(F_i) \, {\rm{dlog}}F \,e^{-\Psi(F_i) {\rm{dlog}}F} \prod_{F_j}\, e^{-\Psi(F_j) {\rm{dlog}}F}} 
\label{eq:L}
\end{equation}

\noindent where $\Psi(F_i)$ is the probability of detecting an object with
line flux between $F$ and $10^{{\rm{d log}}F}F$. This simplifies to Equation \ref{eq:L2}, where {F$_k$} is the product over all bins:

\begin{equation}
{ {\Lambda} = {\prod_{F_i}  \Psi(F_i) \, {\rm{dlog}}F \,  \prod_{F_k}\, e^{-\Psi(F_k) {\rm{dlog}}F}}}
\label{eq:L2}
\end{equation}

We then define the
likelihood function as $S = -2 {\rm{ln}} \Lambda$ and minimize this
(Equation \ref{eq:S}) in order to determine the parameters of the model
($\phi$$^{*}$, $L^{*}$, $\alpha$) that best reproduce the observed data.

\begin{equation}
{ {S} = {-2 \sum }\, {\rm{ln}} \Psi(F_i) + 2 \int \Psi (F)\, {\rm{dlog}}F}
\label{eq:S}
\end{equation}

To obtain $\Psi\,(F)$ (the observed line luminosity function) from
$\phi\,(L)$ (the true line luminosity function) we must consider the
selection effects imposed in observing the data and
statistically model these perturbations to $\phi\,(L)$ before
minimising the likelihood function.

\subsection{Treatment of the filter profile}
To produce a LF, we must first assess the NB filter
response for the emission line of interest as a function of
redshift. A given line flux may arise from a line near the centre of of a
detection filter, or an intrinsically brighter line lying
closer to the edge of the filter where transmission is lower. 

We define $T(z)$, the relative transmission with redshift, to determine the
redshift range to which each filter/line combination is sensitive,
accounting for filter transmission in addition to contributions from
other lines. $T(z)$ is calculated by
propagating a synthetic spectrum through the filter
transmission curve (an example of the resultant line transmission is
seen in Figure \ref{fig:filttrans}), using standard line ratios
([N{\sc~ii}]/H$\alpha$=0.33,  [O{\sc~iii}]/H$\beta$ = 3,  [O{\sc~ii}]
in the low density limit; \citealt{Osterbrock&Ferland89})
normalized to the line of interest. $F$, our line flux, now becomes:

\begin{equation}
{F = \left(\frac{L}{4 \pi d_l^2(z)} \right) T(z)}
\label{eq:T(z)}
\end{equation}

where $L/4 \pi d_l^2(z)$ is the true flux, convolved with
filter transmission $T(z)$. $\Psi(F)$, the probability of seeing a
line of flux $F$, now becomes:

\begin{equation}
{\Psi(F) {\rm{dlog}}F = \int \phi \left(\frac{4 \pi d_l^2(z)F}{T(z)}\right)
  \, \left(\frac{{\rm{d}}V}{{\rm{d}}z}\right)\, {\rm{d}}z\, {\rm{dlog}}L} 
\label{eq:PsiwithT}
\end{equation}

\noindent where ${\rm{d}}V/{\rm{d}}z$ is the co-moving volume interval in our field per unit
redshift, and the integral is performed over the redshift range to
which the filter is sensitive.

\subsection{Line detectability}
\label{section:detfrac}

The detection
of an emission line in our study requires it to reside in an
observable part of parameter space in the NB magnitude --
colour-excess plane. i.e. the region within the three dotted lines in
Figure \ref{fig:trumpetSch}.
The detection
fraction as a function of line flux, is then given by Equation \ref{eq:fdet_dummy}:

\begin{equation}
{f_{{\rm{det}}}(F) = \frac{\int^{m_{\rm faint\ limit}}_{m_{\rm bright\ limit}}\phi_{\rm{BB}}(m_{\rm{BB}}){\rm{d}}m_{\rm{BB}}}{\int^{m_{\rm{faint}}}_{-\infty}\phi_{\rm{BB}}(m_{\rm{BB}}){\rm{d}}m_{\rm{BB}}}}
\label{eq:fdet_dummy}
\end{equation}

\noindent where $\phi_{\rm{BB}}(m_{\rm{BB}})$ is a Schechter
fit to the interpolated broad-band magnitudes of all galaxies
within a particular filter's redshift range, and is derived from the photometric
redshift catalogue of \cite{Grutzbauch11}. The numerator of
${f_{\rm{det}}}$ denotes the sample of galaxies our technique is
capable of detecting (thick red lines in Figure \ref{fig:trumpetSch}), and the denominator denotes the entire
population of galaxies capable of hosting the line down to some
integration limit, ${m_{\rm{faint}}}$. ${m_{\rm{faint}}}$ is the assumed faintest
magnitude of a galaxy capable of hosting a particular emission line
luminosity. Different authors set this limit
differently and somewhat arbitrarily. Here we adopt a limit based on
the rest frame equivalent width of the
line of interest seen in the local Universe, and set EW$_{m_{\rm{faint}}}$$ = 100$\AA\, \citep{Kennicutt92}. The resultant Schechter fit to our observed line fluxes is sensitive to the integration limit of
the denominator (${m_{\rm{faint}}}$). The effect of varying this
limit is discussed in Section \ref{faintend}.

The limits of integration for the numerator are somewhat more
    complex. The detection of a galaxy of particular broad-band
    magnitude for a line of particular line flux
is limited at the bright end by the fainter of 2 broad-band
magnitudes: either that corresponding to the EW limit, or that
corresponding to the bright NB limit. At the faint end, broad-band
magnitude is limited by the brighter of two broad-band
magnitudes: that corresponding to the 5$\sigma$ NB detection limit, or that
corresponding to the faintest broad-band magnitude capable of hosting
a line of this line flux (i.e. ${m_{\rm{faint}}}$). Figure \ref{fig:trumpetSch} illustrates the location of the bright and
faint broad-band limits in relation to lines of constant line flux.

\subsubsection{Broad-band magnitude bright limit}
For the majority of lines the bright limit is set by m$_{\rm{EWlim}}$
(depicted as point A1 in Figure \ref{fig:trumpetSch})
and can be
understood by considering our selection criteria. For an object to be
included in our analysis it must display a colour excess of at
least 3$\sigma$ significance. At bright magnitudes (where
intrinsic colour--scatter dominates uncertainty) detections are
limited by an equivalent width limit (dotted horizontal line in Figure \ref{fig:trumpetSch}). Therefore, for a
given line flux, only host galaxies fainter than a particular broad-band magnitude will result in an equivalent
width large enough to be selected in our sample.

For very bright lines however (such as that traced on the left hand
side of Figure \ref{fig:trumpetSch}) the limiting factor becomes the
saturation limit for point sources in the broad-band imaging used to
determine narrow-band excess. Very
bright lines (of
flux greater than this) will not be selected. The value of this limit
varies from filter to filter, and we define each limit as the point at
which the locus of stellar objects crosses the $3 \sigma$ color--excess
line. Brighter than this limit, emission-line objects could be
indistinguishable from stellar objects. The broad-band saturation limits for each NB filter
are as follows: NB503=19.0, NB570=19.1, NB816=19.25, NB921=19.2,
NB980/NB990=16.0. We note that the effect of this limit is small,
indeed removing this correction from the analysis altogether and
integrating Equation \ref{eq:S} to infinite flux has no
effect above $z \sim 0.6$, and very little effect below this.
An example of a particularly bright line (of 60 $\sigma$ significance) is shown in Figure
\ref{fig:trumpetSch} depicted by the line on the far left. This
demonstrates how only lines of very large line flux are limited by
this saturation limit (point A2 in Figure \ref{fig:trumpetSch}) as
opposed to the EW limit (A1).

\subsubsection{Broad-band magnitude faint limit}
The faint limit for the majority of lines is set by
${m_{\rm{faint}}}$, i.e. the true limiting magnitude of the population
of galaxies capable of hosting that line, as detailed above.

\begin{figure}
\begin{center}
\resizebox {0.48\textwidth}{!}{\includegraphics{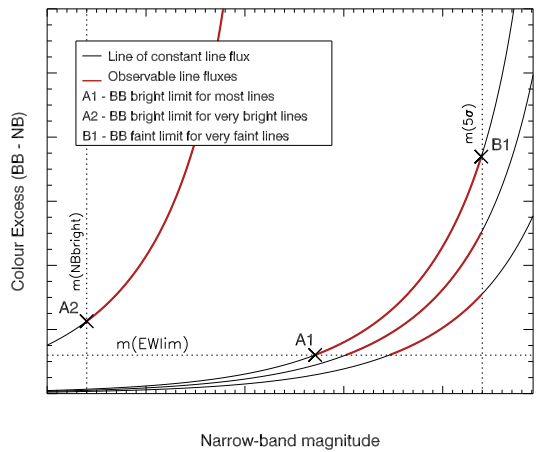}} 
\caption[]{Lines of constant line flux in the NBmag - (colour-excess)
  plane. The dotted lines represent the limiting equivalent width for an
  emission line to meet the selection criteria, and the faintest
  broad-band magnitude that when combined with a given line flux will meet the
  5 $\sigma$ narrow-band threshold. A and B represent points on a line of constant line flux
  that correspond to the bright and faint broad-band limiting
  magnitudes respectively. Thick red lines highlight the portion of
  each line of constant line flux that is detectable in our sample. }
\label{fig:trumpetSch}
\end{center}
\end{figure}

For exceptionally faint lines however, the detection limit of our
catalogue comes into play. For an
object to be detected its broad-band magnitude must be bright enough that when added
to its line flux its NB magnitude is above the 5$\sigma$ detection
threshold, and so for very faint lines this constraint sets the faint
limit on broad-band magnitude. Point B1 represents this limit for a
particular line flux in
Figure \ref{fig:trumpetSch}.

Equation \ref{eq:fdet_dummy} then becomes Equation \ref{eq:fdet}:

\begin{equation}
{f_{{\rm{det}}}(F) = \frac{\int^{\rm min(m_{5\sigma},m_{\rm
        faint})}_{\rm max(m_{\rm{EWlim}} ,m_{\rm NB bright})}\phi_{\rm{BB}}(m_{\rm{BB}}){\rm{d}}m_{\rm{BB}}}{\int^{m_{\rm{faint}}}_{-\infty}\phi_{\rm{BB}}(m_{\rm{BB}}){\rm{d}}m_{\rm{BB}}}}
\label{eq:fdet}
\end{equation}

\noindent and $\Psi(F)$ now takes its
final form:

\begin{equation}
{\Psi(F)\, {\rm{d log}} \,F=f_{\rm{det}}\,(F) \int \phi
  \left(\frac{4\pi
      d^{2}_{l}F}{T(z)}\right)\frac{{\rm{d}}V}{{\rm{d}}z}\,{\rm{d}}z\,{\rm{d log}} \,F.}
\label{eq:LF}
\end{equation}

We note that the presence of a strong correlation between SFR and
stellar mass (e.g. \citealt{Noeske08}, however see
\citealt{SobralMass11} for an alternative view) would mean that low--luminosity
lines would be less likely to be found in more massive galaxies. We
will address this issue and its effect on line detectability in paper
2, where we will study luminosity functions as a function of stellar
mass. 

\subsection{Detection Completeness}
Finally, we assess the detection completeness of the narrow-band
selection catalogue retrieved via \textsc{SExtractor}.
The completeness of detections is dependent
on the depth of the detection image and on the size of the aperture
used. As our analysis is carried out in 2 arcsecond apertures and a retrospective
aperture correction applied (see below), we assess completeness in the
same manner down to the 5$\sigma$ limiting magnitude. We construct a
series of images inserting randomly positioned fake point sources
in each image, allowing us to determine recovery fraction as a function of magnitude. We estimate
incompleteness per $\Delta$m$_{{\rm{NB}}}=0.05$ bin, and create an
appropriate number of sources with NB
excesses distributed following the observed distribution of detected
objects, and with NB magnitudes uniformly distributed within the bin. The
effect is small as we only populate the faintest bins,
typically adding fewer than 10 objects to a LF. This number rises to
21 objects (a 2 per cent increase) for the second most densely populated LF ([O{\sc~ii}] in
NB816) and 91 objects (a 4 per cent increase) for [O{\sc~ii}] in NB921 where the number of
detections is more than a factor of 2 larger again. Corrections are
particularly low due to the stringent $5\sigma$ NB-detection
requirement.

\subsection{Galactic Extinction and Aperture Correction}
\label{Gal_Aper}
Each LF is corrected for Galactic extinction of $A_V
=0.070$ magnitudes using the \cite{CardelliClaytonMathis}
reddening law (following the same method as in Section \ref{sect:phot}). As the
analysis is carried out using aperture photometry, we must
assess the fraction of flux that is lost by only considering the light
within our 2 arcsecond aperture. We calculate the ratio of total NB
flux to aperture NB flux using the output \textsc{SExtractor} values
(\textsc{flux\_best} and \textsc{flux\_aper} respectively) for each
redshift slice, and scale each output
LF according to the median ratio in that slice. The correction is typically small (a factor of $\sim$
1.2) even at the lowest redshifts, however for the smoothed VISTA NB
data we find a slightly larger correction ($\sim$ 1.8) due to a much
broader PSF.

\section{Results}
\label{results}

We derive luminosity functions to examine the evolution of the H$\alpha$,  [O{\sc~iii}] and
[O{\sc~ii}] luminosity functions. The resultant Schechter fits provide LFs in 12 redshift slices between
0.14 and 1.64. Using the parametrizations of these LFs we can derive
SFRs and compute $\rho_{\rm{SFR}}$ to examine the evolution of
star formation with redshift.

\subsection{Luminosity Functions}
\label{LFs}
The derived luminosity functions are presented in Table \ref{tab:LF}
and Figure \ref{fig:LF}. Volumes quoted are representative of the volume covered
where filter transmission is greater than 0.5. Observed (aperture-corrected) values are presented,
followed by the extinction-corrected value of $L^*$, assuming 1 magnitude of
extinction at H$\alpha$ (\citealt{Kennicutt92}; 1998) and as applied in
\cite{Fujita03}, \cite{Pascual05},
\cite{Sobral12_double} and \cite{Sobral12}. Figure \ref{fig:LF} presents the best fitting LFs for the data at each
redshift with the maximum likelihood fit shown in bold coloured
lines. Symbols represent the
  binned luminosity function in bins of $\sim$0.1 dex.

Alongside each panel of LFs, we include a graphic representation of
the error ellipses of the 1, 2 and 3$\sigma$ contours in $L^*$-$\phi^*$
space representing the boundaries of the volume in 2D ($\phi^*$,
$L^*$) space. We anchor $\alpha$ to the median value across
the redshift slices for
each line as shown in each Figure, while $L^*$ and $\phi^*$ are allowed to vary (i.e. $\alpha$
is constant per panel to allow comparison of LFs). 

\subsubsection{Treatment of Shallow Data}
\label{shit}
As discussed in Section \ref{ML}, the maximum likelihood analysis
minimizes Equation \ref{eq:S} allowing $\phi^*$, $L^*$
and $\alpha$ to vary. However, this analysis requires a deep and
plentiful data set in order to produce a well constrained fit,
particularly for $\alpha$. NB570 is the
shallowest of the Subaru NB images, and due to its blue wavelength
samples [O{\sc~iii}] emitters at a very low redshift (in the smallest
cosmic volume of the analysis); consequently the LFs produced from the
NB570 filter selection ([O{\sc~iii}] at z=0.14 and [O{\sc~ii}] at 0.53) are subject
to low number statistics, with only 42 and 83 objects respectively
making up each LF. 

Furthermore, VISTA data are $\sim$1 mag shallower than NB570, and so our data
struggles to probe $L^{*}$ in the $z=0.5, 1.0$ and $1.6$ redshift
slices, in addition to very low numbers of emitters being detected. 

The shallower depth of this data means that $\alpha$ is not as well
constrained as for the deeper more detection-rich LFs, however,
the NB985 combined data set is selected close in
wavelength to the NB921 filter, and consequently probes redshift
slices only marginally more distant than the NB921 selection. To
combat the shallow data we fit LFs for the 3 VISTA redshift slices
with $\alpha$ constrained by a Gaussian prior based on the value
for the NB921 selection for the same emission line. This is a valid method of
evaluation where we do indeed have some prior knowledge of the value
of $\alpha$. We take a similar approach with the NB570 [O{\sc~ii}] LF at
$z=0.53$, setting
the prior according to the [O{\sc~ii}] LF at $z=0.35$.

\subsubsection{The Effect of Cosmic Variance}
\label{cos_var}
\begin{figure*}
\begin{center}
\resizebox{0.48\textwidth}{!}{\includegraphics{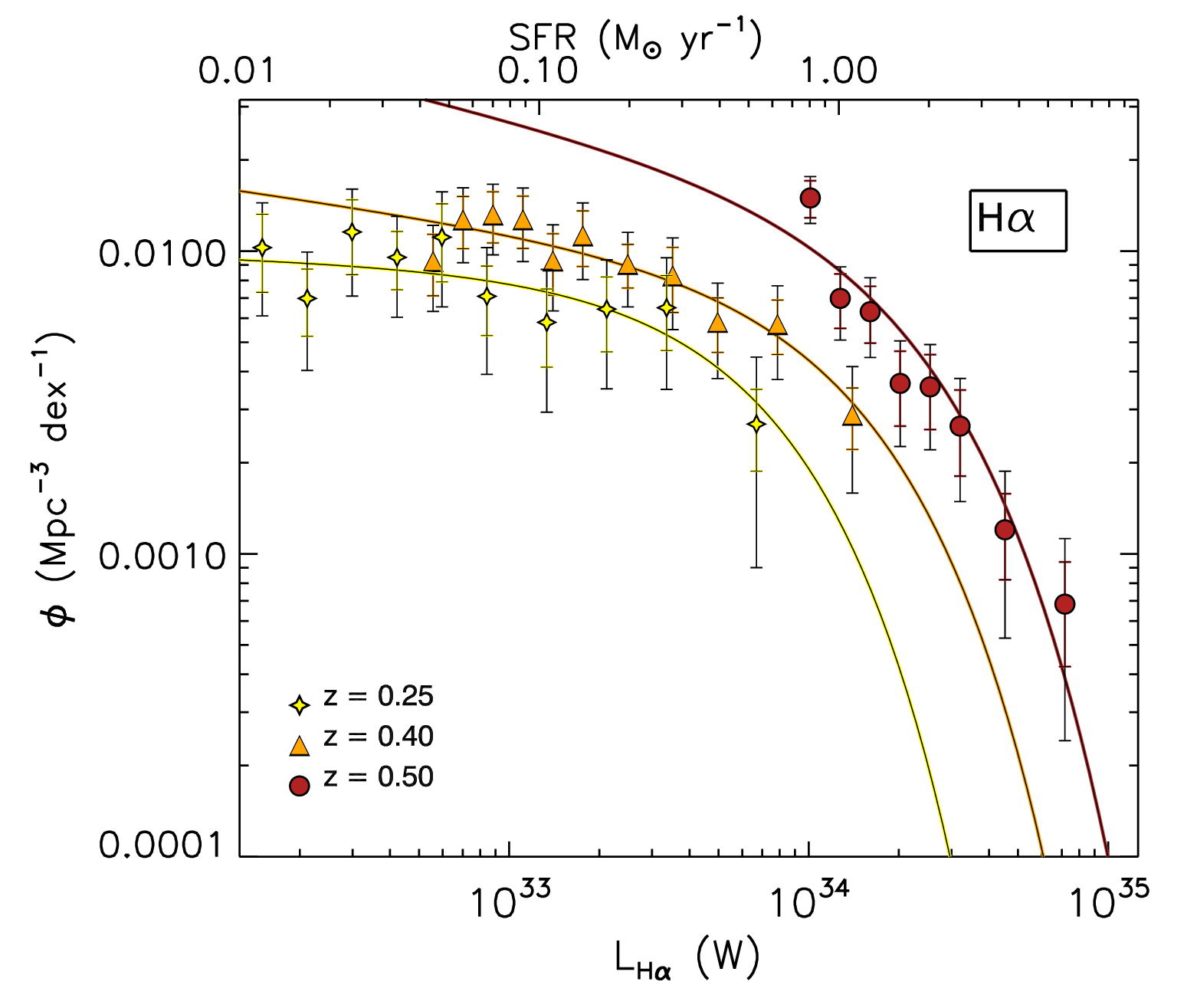}}
\resizebox{0.48\textwidth}{!}{\includegraphics{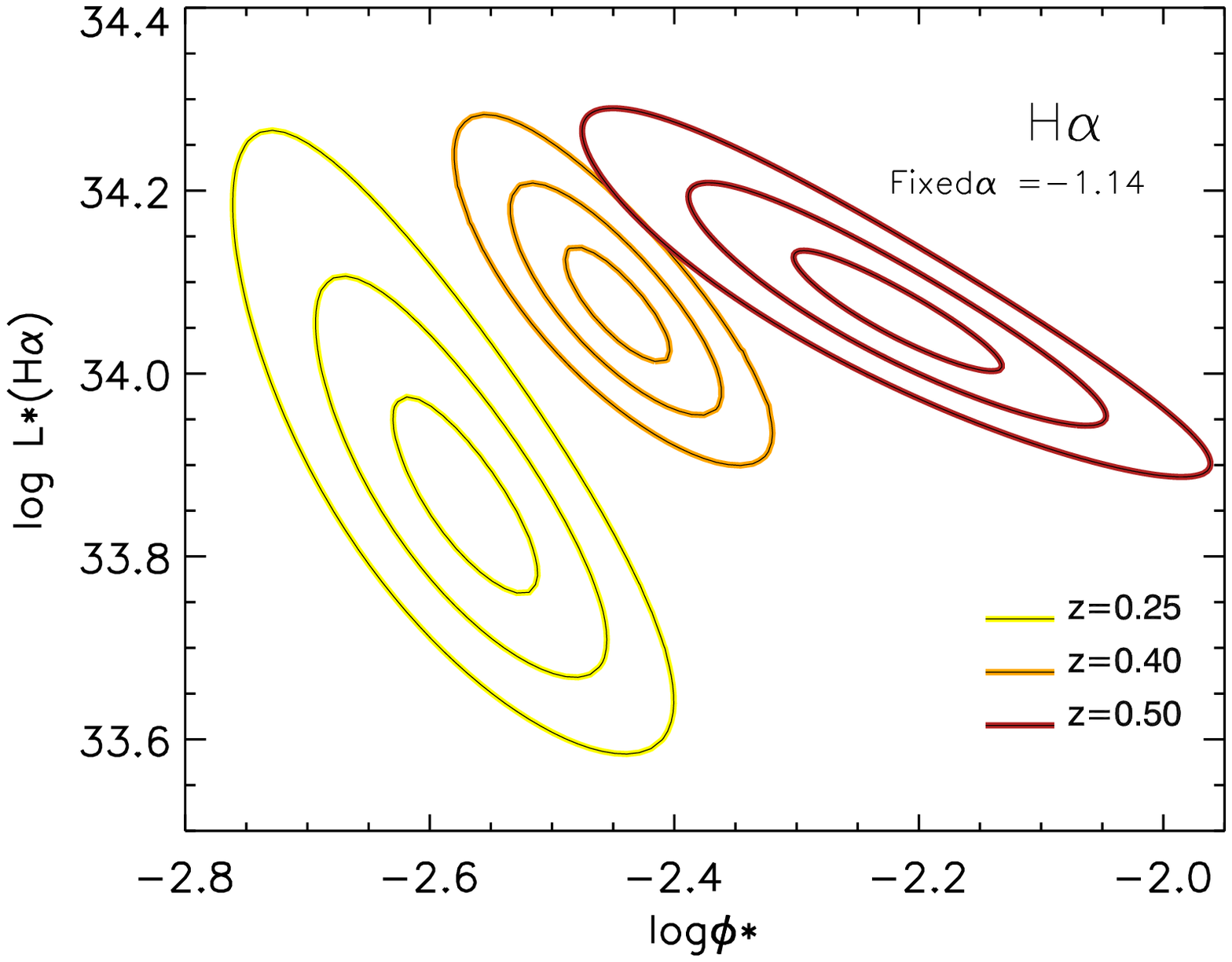}}
\resizebox{0.48\textwidth}{!}{\includegraphics{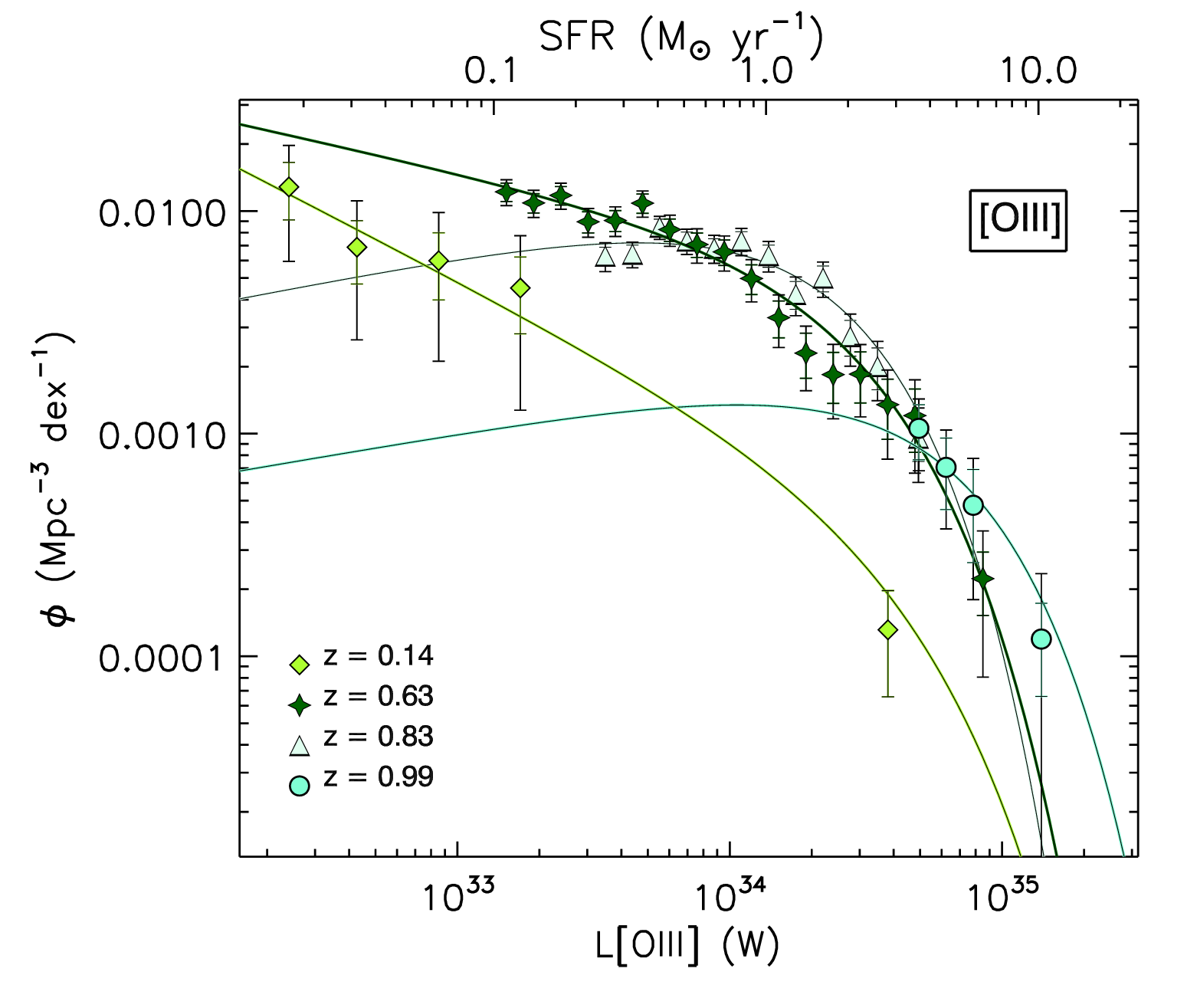}}
\resizebox{0.48\textwidth}{!}{\includegraphics{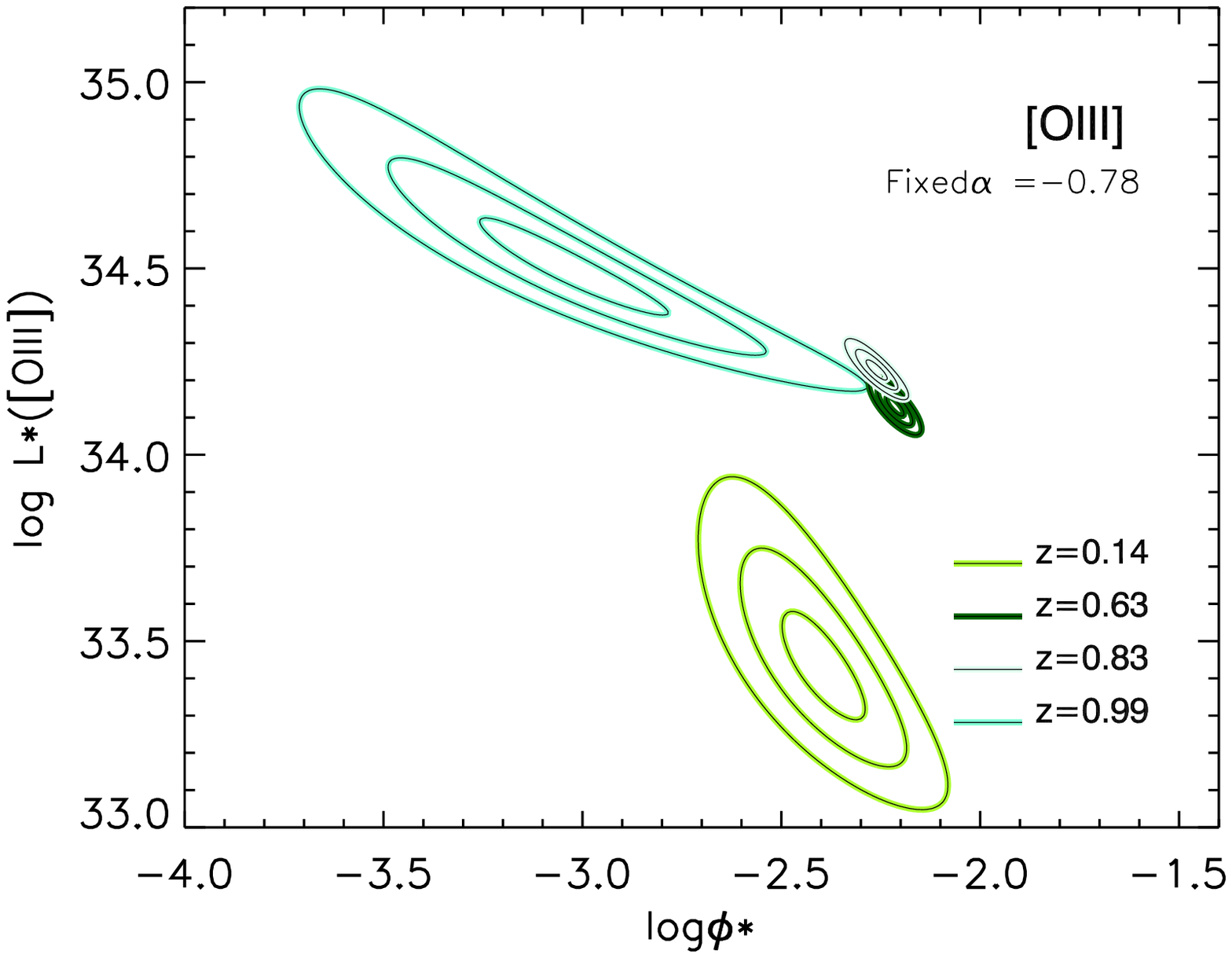}}
\resizebox{0.48\textwidth}{!}{\includegraphics{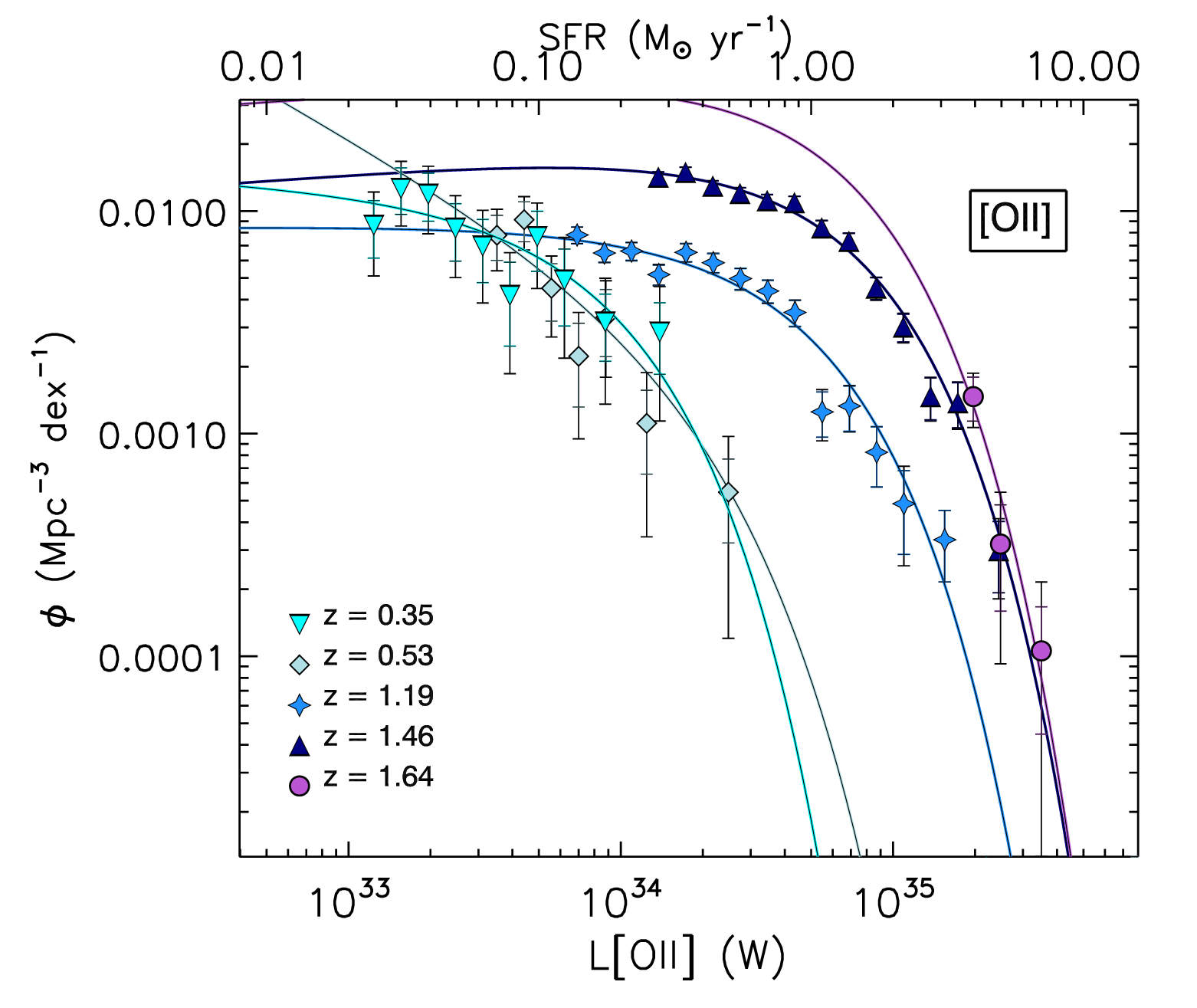}}
\resizebox{0.48\textwidth}{!}{\includegraphics{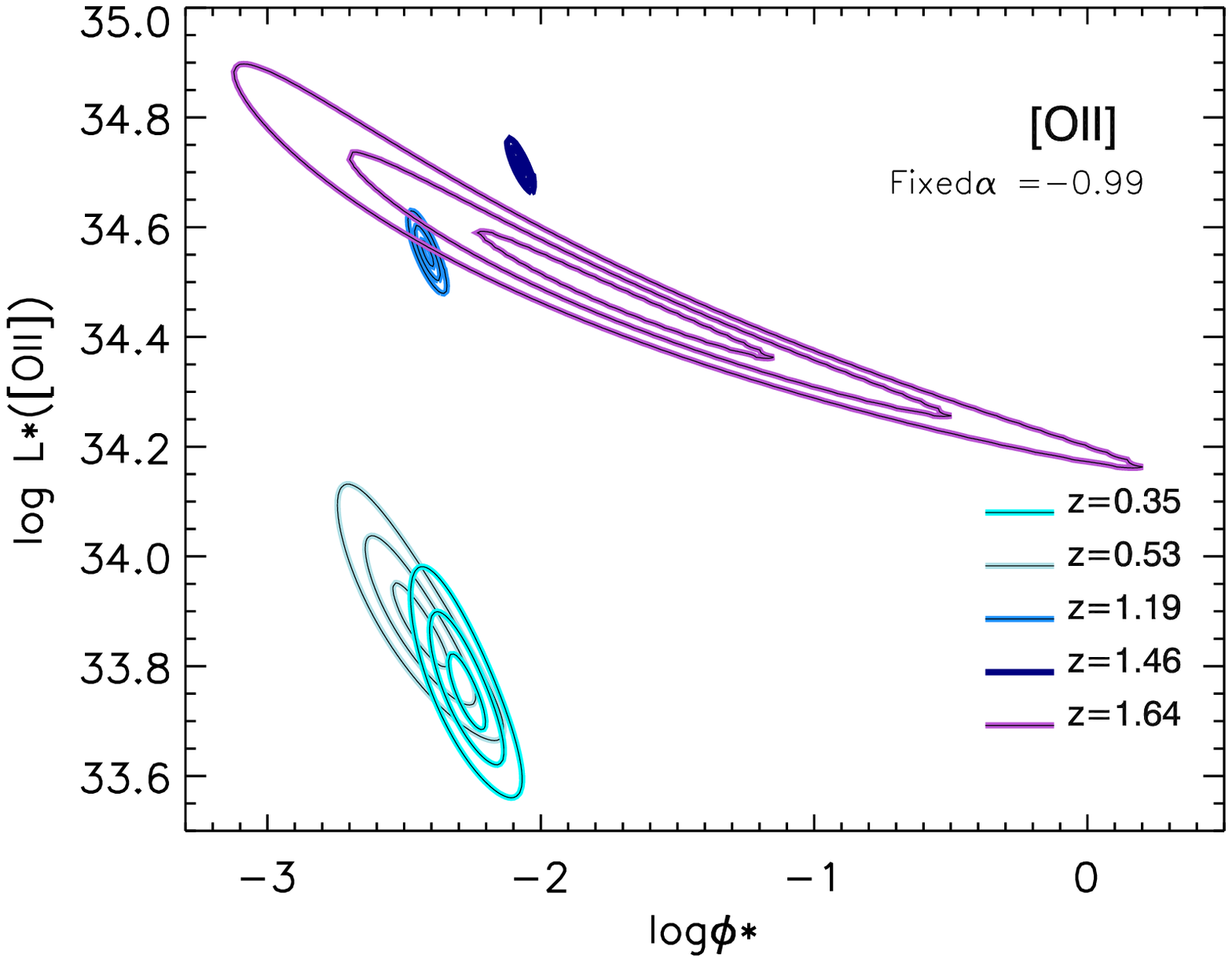}}

\caption[]{Graphic representation of our best--fitting luminosity
  functions (Gaussian prior on $\alpha$ for VISTA LFs and NB570 at
  $z=0.53$) and
  their 1, 2 and 3$\sigma$ error ellipses ($\alpha$ fixed to the
  median value per line for comparison) in $L^*$-$\phi^*$
  space. The smaller of the two error bars on each point in the lefthand panels are Poissonian,
  larger error bars have the 1$\sigma$ variation as estimated by our
  jackknifing technique added in quadrature. H$\alpha$ (top panels)
  [O{\sc~iii}] (central panels) and [O{\sc~ii}] (lower panels). The
  upper SFR scale on these plots gives an indication of the raw SFR
  values for these LFs, these numbers have not been corrected for dust
extinction.}
\label{fig:LF}
\end{center}
\end{figure*}

In order to study the effect of our limited number of detections we
undertake a jackknife analysis in every redshift slice. For each slice
we use random sampling with replacement to construct 1000 synthetic
catalogues of narrow-band emitters such that each sample is of the same size as our
true catalogues. We fit luminosity functions to each of these synthetic
catalogues and measure the standard deviation in each parameter. These
$1 \sigma$ deviations are comparable to but no bigger than the
uncertainties in the true dataset. Errors quoted in Table
\ref{tab:LF} refer only to the $1\sigma$ uncertainty in the real
data. In Figure \ref{fig:LF} we plot two error bars for each
point. The smaller of the two error bars represents the Poissonian
error, and the larger includes the standard
deviation on the number of objects in
each luminosity bin from the jackknifed samples added in quadrature.

\begin{table*}
\caption{Luminosity functions in 12 redshift slices for H$\alpha$,
  [O{\sc~iii}] or [O{\sc~ii}] emission. Numbers of objects quoted
  refer to the final number of objects per luminosity function (after
  culling to the broadest 3$\sigma$ colour-scatter limit of the SXDS
  fields, and taking into account the
  completeness correction). Volumes refer to the volume enclosed where the
  filter  transmission is greater than 50 per cent. Schechter fits to the data are
  presented with the value of L$^*$ corrected for aperture effects and
  Galactic extinction at the wavelength of the filter. The extinction
  corrected values listed are those derived for 1 mag of extinction at
  H$\alpha$. Where the raw data did not produce a converging LF, values are absent for SFR$^{*}$ and
  $\rho_{\rm{SFR}}$ as these would be
  non-physical.}
\label{tab:LF}
\begin{centering}
\begin{tabular}{ccccccccc} \hline
&&& & && \\ 
&&& & &&\multicolumn{3}{c}{\bf{Extinction - corrected}}  \\
\cline{7-9}
%&&& & && && \\ 
{\bf{Redshift}}   & {\bf{Objects}}  & {\bf{Volume} } & 
{\bf{log $\phi^{*}_{\rm{H}_{\alpha}}$}}  & 
{\bf{log L$^{*}_{\rm{H}_{\alpha}}$}} & 
{\bf{$\alpha_{\rm{H}_{\alpha}}$}} &  
{\bf{log L$^{*}_{\rm{H}_{\alpha}}$}} & 
{\bf{SFR$^{*}$}} &
{\bf{$\rho_{\rm{SFR}_{\rm{H}_{\alpha}}}$}} \\

& & (10$^4$ Mpc$^3$) &
(Mpc$^{-3}$) & (Watts) & 
& (Watts) & (M$_{\odot}$ yr$^{-1}$) & (M$_{\odot}$ yr$^{-1}$ Mpc$^{-3}$)\\
\hline 
 {\bf{0.25}} & 142 & 1.22 & -2.43$^{+0.17}_{-0.21}$ & 33.83$^{+0.19}_{-0.16}$ &
 -1.03$^{+0.17}_{-0.15}$ & 34.23$^{+0.19}_{-0.16}$ &
 1.33$^{+0.78}_{-0.41}$ & 0.0051$^{+0.0146}_{-0.0008}$ \\
 {\bf{0.4}} & 271 & 2.95 & -2.44$^{+0.14}_{-0.17}$ & 34.16$^{+0.13}_{-0.11}$ &
  -1.14$^{+0.14}_{-0.13}$ & 34.55$^{+0.13}_{-0.11}$ & 2.85$^{+0.99}_{-0.64}$ & 0.0113$^{+0.0005}_{-0.0005}$ \\  
  {\bf{0.5}}& 151 & 4.76 & -3.27$^{+0.55}_{-0.94}$ & 34.69$^{+0.46}_{-0.27}$ &
  -2.16$^{+0.33}_{-0.31}$ &--- &  --- & ---\\  
 {\bf{0.5}}$^{\dagger}$&  151 & 4.76 & -2.23$^{+0.10}_{-0.12}$ & 34.34$^{+0.09}_{-0.07}$ &
  -1.23$^{+0.12}_{-0.13}$ & 34.74$^{+0.09}_{-0.07}$ &
  4.33$^{+1.00}_{-0.73}$ &0.0308$^{+0.0032}_{-0.0022}$\\ 
\hline

{\bf{Redshift}}   & {\bf{Objects}}  & {\bf{Volume} } & 
{\bf{log $\phi^{*}_{\rm[OIII]}$}}  & 
{\bf{log L$^{*}_{\rm[OIII]}$}} & 
{\bf{$\alpha_{\rm[OIII]}$}} &  
{\bf{log L$^{*}_{\rm[OIII]}$}} & 
{\bf{SFR$^{*}$}} &
{\bf{$\rho_{\rm{SFR}_{\rm[OIII]}}$}} \\

& & (10$^4$ Mpc$^3$) &
(Mpc$^{-3}$) & (Watts) & 
& (Watts) & (M$_{\odot}$ yr$^{-1}$) & (M$_{\odot}$ yr$^{-1}$ Mpc$^{-3}$)\\
\hline 

 {\bf{0.14}} &  42 & 0.36 & -3.67$^{+1.11}_{-\infty}$ & 34.60$^{+\infty}_{-0.96}$ &
  -1.63$^{+0.60}_{-0.24}$ & 35.14$^{+\infty}_{-0.96}$& 10.17$^{+\infty}_{-9.05}$ & 0.0040$^{+\infty}_{-0.0008}$\\
 {\bf{0.63}} & 943 & 8.09 & -2.57$^{+0.11}_{-0.13}$ & 34.44$^{+0.09}_{-0.08}$ &
  -1.27$^{+0.11}_{-0.11}$ &34.99$^{+0.09}_{-0.08}$ & 7.16$^{+1.65}_{-1.20}$ & 0.0240$^{+0.0038}_{-0.0025}$\\       
{\bf{0.83}} & 910 & 12.35 & -2.25$^{+0.07}_{-0.09}$ & 34.28$^{+0.07}_{-0.08}$ &
  -0.76$^{+0.22}_{-0.19}$ & 34.83$^{+0.07}_{-0.08}$ &
  4.95$^{+0.87}_{-0.83}$ & 0.0252$^{+0.0780}_{-0.0084}$\\    
{\bf{0.99}}& 32 & 14.69 & -14.45$^{+\infty}_{-\infty}$ & 38.65$^{+\infty}_{-\infty}$ &
  -3.78$^{+0.36}_{-0.40}$ & --- &  --- & --- \\     
{\bf{0.99}}$^{\dagger}$ & 32 & 14.69 & -3.00$^{+0.23}_{-0.22}$
& 34.70$^{+0.14}_{-0.12}$ & -0.78$^{+0.19}_{-0.21}$ &
35.25$^{+0.14}_{-0.12}$  & 12.97$^{+6.22}_{-3.57}$ & 0.0119$^{+0.0035}_{-0.0024}$\\  
\hline

{\bf{Redshift}}   & {\bf{Objects}}  & {\bf{Volume} } & 
{\bf{log $\phi^{*}_{\rm[OII]}$}}  & 
{\bf{log L$^{*}_{\rm[OII]}$}} & 
{\bf{$\alpha_{\rm[OII]}$}} &  
{\bf{log L$^{*}_{\rm[OII]}$}} & 
{\bf{SFR$^{*}$}} &
{\bf{$\rho_{\rm{SFR}_{\rm[OII]}}$}} \\

& & (10$^4$ Mpc$^3$) &
(Mpc$^{-3}$) & (Watts) & 
& (Watts) & (M$_{\odot}$ yr$^{-1}$) & (M$_{\odot}$ yr$^{-1}$ Mpc$^{-3}$)\\
\hline 
{\bf{0.35}} & 112 & 2.22 & -2.31$^{+0.19}_{-0.29}$ & 33.90$^{+0.20}_{-0.16}$ &
 -1.06$^{+0.38}_{-0.34}$ & 34.65$^{+0.20}_{-0.16}$  &
 1.11$^{+0.65}_{-0.34}$ & 0.0142$^{+0.0010}_{-0.0007}$\\  
{\bf{0.53}} & 83 & 3.95 & -5.55$^{+1.75}_{-\infty}$ & 35.38$^{+\infty}_{-0.84}$ &
  -2.70$^{+0.36}_{-0.17}$ & --- & ---  & --- \\ 
{\bf{0.53}}$^{\dagger}$ & 83 & 3.95 & -2.85$^{+0.28}_{-0.41}$ & 34.13$^{+0.22}_{-0.17}$ &
 -1.68$^{+0.33}_{-0.31}$ & 34.98$^{+0.22}_{-0.17}$ & 2.34$^{+1.54}_{-0.76}$  & 0.0231$^{+0.0080}_{-0.0052}$\\  
{\bf{1.19}} & 981 & 19.06 & -2.41$^{+0.07}_{-0.09}$ & 34.61$^{+0.07}_{-0.06}$ &
 -0.99$^{+0.14}_{-0.13}$ & 35.37$^{+0.06}_{-0.06}$ & 5.76$^{+1.01}_{-0.74}$ & 0.0557$^{+0.0025}_{-0.0019}$\\        
{\bf{1.46}} & 2218 & 23.09 & -2.03$^{+0.04}_{-0.06}$ & 34.76$^{+0.05}_{-0.04}$ &
 -0.91$^{+0.11}_{-0.11}$ & 35.52$^{+0.05}_{-0.04}$ & 8.09$^{+ 0.99}_{-0.71}$ & 0.1796$^{+0.0053}_{-0.0055}$\\     
{\bf{1.64}}& 27 & 28.31 & -2.98$^{+0.79}_{-\infty}$ & 35.27$^{+\infty}_{-0.31}$ &
  -3.10$^{+1.35}_{-1.15}$ & --- &  ---  & --- \\ 
{\bf{1.64}}$^{\dagger}$ & 27 & 28.31 & -1.68$^{+0.50}_{-0.43}$ &
34.73$^{+0.10}_{-0.11}$ & -0.91$^{+0.11}_{-0.10}$ & 35.49$^{+0.10}_{-0.11}$ & 7.47$^{+1.93}_{-1.67}$ & 0.3711$^{+0.5775}_{-0.1951}$\\      
\hline
\multicolumn{9}{l}{$\dagger$ A Gaussian prior on $\alpha$ is applied
  to these luminosity functions.}
\end{tabular}
\end{centering}
\end{table*} 

\subsection{Star Formation Rates}
We use the best-fitting Schechter function in each of the 12 redshift
slices to calculate the characteristic SFR and $\rho_{\rm{SFR}}$ across
10 Gyr of cosmic time. SFRs are calculated according to the relations
of \cite{Kennicutt98} for H$\alpha$ (Equation \ref{eq:SFRHa}), [O{\sc~ii}]  (Equation
\ref{eq:SFROII}) and using the standard line ratios:
H$\alpha$/H$\beta$=2.78 and [OIII]/H$\beta$=3 \citep{Osterbrock&Ferland89} to derive
Equation \ref{eq:SFROIII} for [O{\sc~iii}]. Equation
\ref{eq:SFROIII} is in good agreement with the empirical relationship
of \cite{Ly07}. This method determines
extinction-corrected SFRs assuming continuous star formation, Case B
recombination and a Salpeter IMF.

\begin{equation}
{{\rm{SFR}}({\rm{M}}_{\odot}{\rm{yr}}^{-1}) = 7.9 \times 10^{-35}
  {{\rm{L_{H \alpha}}} {{\rm{E}(H \alpha)}}}}%{\rm{W}}}}
\label{eq:SFRHa}
\end{equation}

\begin{equation}
{{\rm{SFR}}({\rm{M}}_{\odot}{\rm{yr}}^{-1}) = 7.35 \times
  10^{-35}{{\rm{L_{[OIII]}}}} {{\rm{E}[OIII]}}} %\times{\rm{W}}}}
\label{eq:SFROIII}
\end{equation}

\begin{equation}
{{\rm{SFR}}({\rm{M}}_{\odot}{\rm{yr}}^{-1}) = 1.39 \times
  10^{-34}{{\rm{L_{[OII]}}}} {{\rm{E}(H \alpha)}}}%\times{\rm{W}}}}
\label{eq:SFROII}
\end{equation}

\begin{figure*}
\begin{center}
\resizebox{0.48\textwidth}{!}{\includegraphics{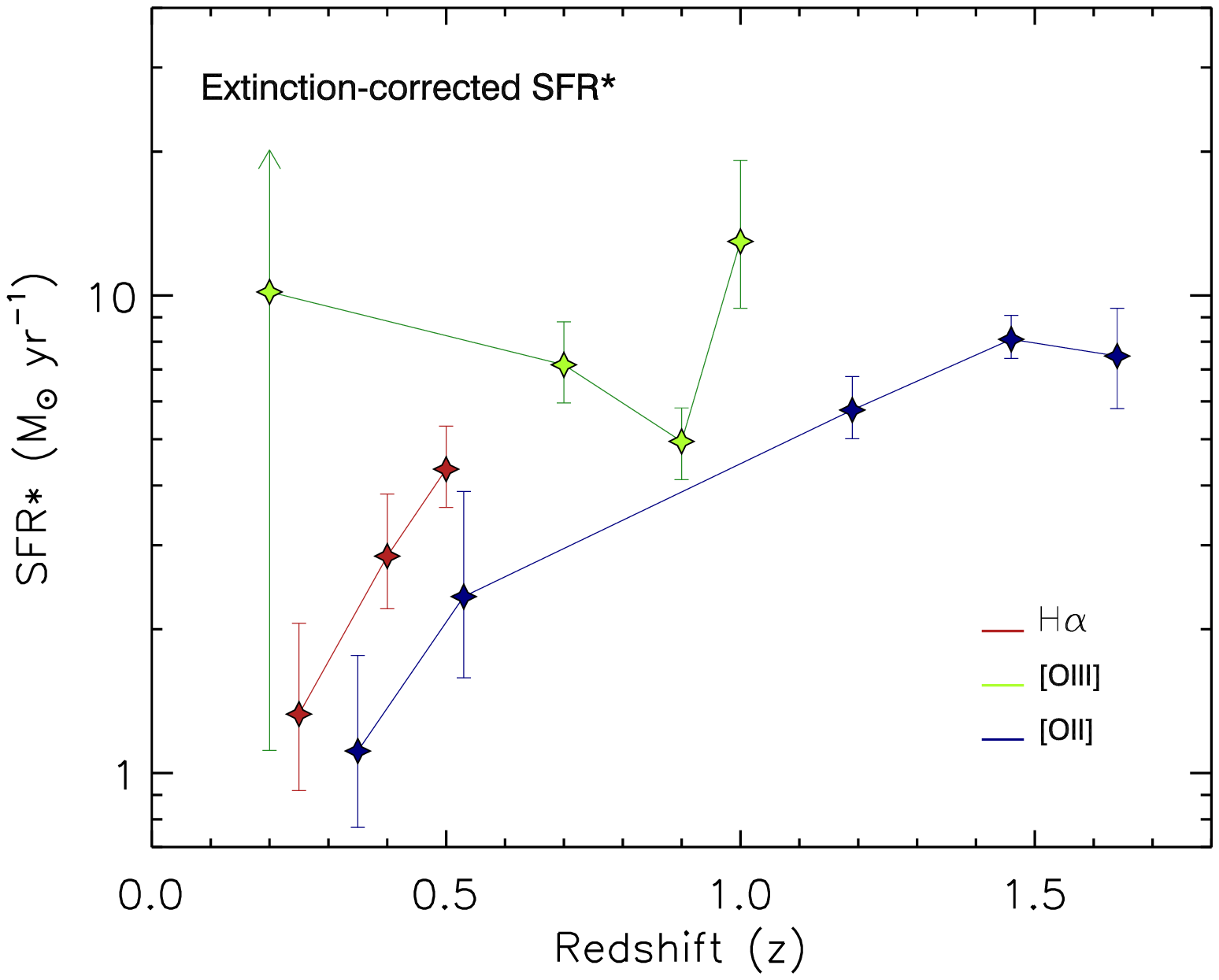}}
\resizebox{0.48\textwidth}{!}{\includegraphics{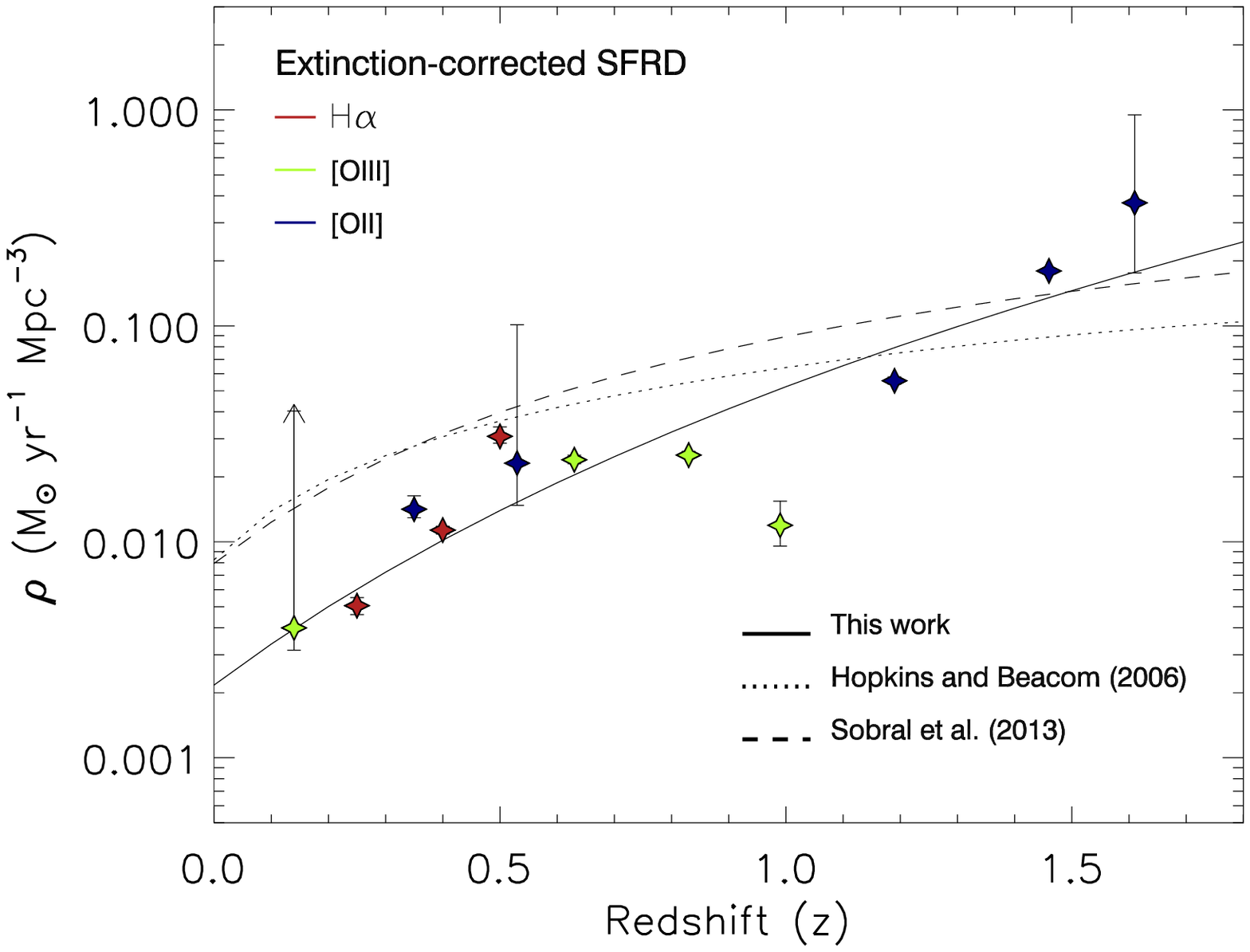}}
\makebox[\textwidth]{\bf (a) \hspace{0.45\textwidth} \bf (b)}
\caption[]{Panel (a) shows the extinction-corrected characteristic
  star formation rate (SFR $^*$) determined in each redshift slice, error bars corresponding
  to the $1\sigma$ error on $L^*$. Colours reference the
  emission line used to derive the SFR. Panel (b) presents
  $\rho$\ $_{\rm{SFR}}$ in each redshift slice. Colours again
represent emission line used to derive the measurement. The solid
black line shows our parametrization of $\rho$\ $_{\rm{SFR}} \propto (1+z)^{4.58}$, and the dotted
lines show parametrisations in the literature.}
\label{fig:SFR}
\end{center}
\end{figure*}

Figure \ref{fig:SFR}\,(a) presents the
fully-corrected characteristic star formation rate (SFR*; aperture and extinction corrections described in Sections \ref{LFs} and
\ref{Gal_Aper}) for each redshift slice in this analysis,
red points showing H$\alpha$ derived estimates, green, [O{\sc~iii}],
and blue, [O{\sc~ii}]. Error bars represent SFRs for the corresponding
1$\sigma$ upper and lower bounds on the characteristic SFR. Figure \ref{fig:SFR}(b) presents the
integrated $\rho_{\rm{SFR}}$ for each LF (full integration,
extinction-corrected) together with the fits of \cite{Sobral12}
and \cite{Hopkins06} using the data of \cite{BaldryAndGlazebrook03}.

\section{Discussion}
\label{discussion}

\subsection{Evolution of the Star Formation Rate}
We find characteristic SFRs broadly consistent between all three
indicators out to $z\sim1.6$. The characteristic SFR increases by a factor $\sim10$ between
$z\sim0.14$ and $z\sim1.6$, in good
agreement with literature results. The H$\alpha$ and [O{\sc~ii}]
derived star formation rates show a distinct increase with redshift consistent
with other studies, but we note that the [O{\sc~iii}] estimates (while
broadly consistent with the other indicators due to their uncertainty)
would produce a more convincing increase with redshift if for instance
the [O{\sc~iii}]/H$\beta$ ratio were to decrease with redshift.

The parametrization of $\rho_{\rm{SFR}} $
according to $\rho_{\rm{SFR}} \propto$ (1+z)$^{\gamma}$ gives
$\gamma$ = 4.58 when handling errors as the maximum and minimum values
of $\rho_{\rm{SFR}}$ produced from the $1\sigma$ confidence limits of
each measurement of $\phi^*$, $L^*$ and $\alpha$. We note
that applying the same least squares fit to
the data of \cite{Ly07} gives a slope of $\gamma$ = 4.20. Applying a
blanket 10 per cent error on $\rho_{\rm{SFR}} $ to our data results in
$\gamma$ = 3.02, similar to the parametrization of \cite{Sobral12}. The
discrepancy between these results and shallower slopes in the
literature is attributed to the small volumes probed in our low
redshift slices. For instance, using the figures of \cite{Somerville04} we
expect uncertainty in $\rho_{\rm{SFR}}$ in the lowest redshift slices to be comparable to the measurements
themselves due to cosmic variance. Larger areas of sky are required for this technique to
be of use in determining the $\rho_{\rm{SFR}}$ at low redshift as the redshift
slices sampled in narrow-band surveys are especially
susceptible to cosmic variance. We return to this point in Section \ref{comp}. We note also that the [O{\sc~ii}] LF of
\cite{Gilbank10} at $z\sim 1$ gives a $\rho_{\rm{SFR}}$ in good
agreement with the progression of [O{\sc~ii}] values seen
in panel \ref{fig:SFR}(b). 

This technique is however sensitive to
particularly low SFRs, probing well below SFR$^*$ ($\ll1 M_{\odot}\,{\rm{yr^{-1}}}$)
in the deepest data sets
allowing a more accurate evaluation of $\alpha$. While the role of these faint
galaxies' contributions to the
overall $\rho_{\rm{SFR}}$ is unclear, if we wish to understand the nature of this
faint population it is important
to first constrain $\alpha$.

\subsection{The Faint End Slope}
\label{faintend}
The gradient of the faint end slope, $\alpha$, is the subject of much
debate. The recent work of \cite{Sobral12} made a significant advance
on previous studies by examining estimates of $\alpha$ across a  broad
redshift range using the first sample of consistently
selected emission line objects out to $z=2.23$. 
An important part of determining the most likely luminosity function
at any redshift however, is to determine the significance of the galaxies we are capable
of detecting in the analysis within the
general population at redshift $z$. As discussed in
Section \ref{section:detfrac}, we compute the ratio of detectable
galaxies to the total number of galaxies capable of hosting the
line. Investigating this ratio we find that the best fit $\alpha$ of
each LF is in fact very sensitive
to the assumed faintest broad-band magnitude of a galaxy that can
host an emission line (i.e. the denominator of this
fraction). The analysis of \cite{Sobral12} assesses emission-line
detection fraction down to the limit of their NB data, and so we
investigate the effect of assessing our data in the same way. 

Evaluating our data set with ${m_{\rm{faint}}}$ set to the limiting NB
magnitude (${m_{\rm{faint}}}$(NB)) we find that the resultant
values of $\alpha$ are consistently steeper than for our ${m_{\rm{faint}}}$(EW)
analysis, the medians of the two samples being offset
by 0.3. Furthermore, the range in values of $\alpha$ 
is reduced (from $0.90$ to $0.66$) through use
of the (${m_{\rm{faint}}}$(NB)) limit across the same redshift range.

\begin{figure}
\begin{center}
\resizebox {0.48\textwidth}{!}{\includegraphics{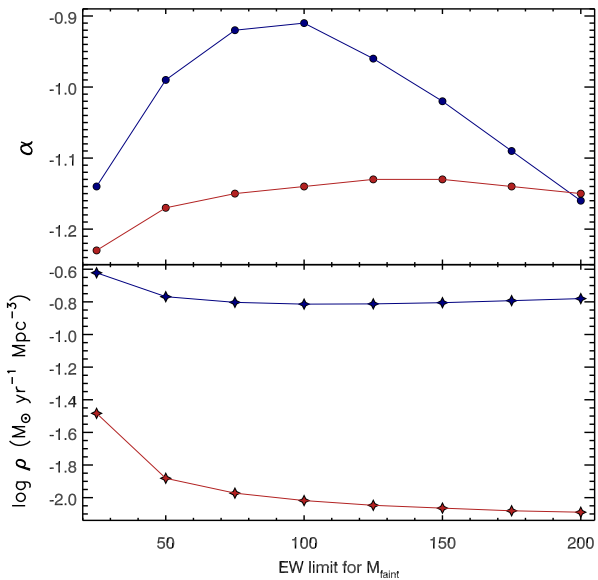}} 
\caption[]{The variation of $\alpha$ and $\rho_{\rm{SFR}}$ with differing input values
of the limiting equivalent width for ${m_{\rm{faint}}}$. The upper panel shows the variation in
resultant values of $\alpha$ for the NB921 [O{\sc~ii}] luminosity
function at $z=1.46$ (blue) and the NB921 H$\alpha$ luminosity function
at $z=0.40$ (red) when the limiting value of equivalent width for
${m_{\rm{faint}}}$ is varied. The lower panel shows
the corresponding resultant values of $\rho_{\rm{SFR}}$ as ${m_{\rm{faint}}}$ changes for
the same two luminosity functions.}
\label{fig:EWonalpha}
\end{center}
\end{figure}

Taking the well-constrained NB921 H$\alpha$ and NB921 [O{\sc~ii}]
luminosity functions we vary the input limiting equivalent width value
for ${m_{\rm{faint}}}$. The sensitivity of $\alpha$ to the input value is displayed in Figure
\ref{fig:EWonalpha}. As EW$_{m_{\rm{faint}}}$ is increased, a larger
fraction of the population of galaxies can host emission lines, and so
the detection fraction decreases and the resultant $\alpha$
steepens. Conversely however, as the input EW$_{m_{\rm{faint}}}$ is
lowered below 100\AA\ the numerator of Equation \ref{eq:fdet} (the red
portion of a line in Figure \ref{fig:trumpetSch}) begins
to decrease when ${m_{\rm{faint}}}$ crosses the 5$\sigma$ detection
limit. As the detection limit moves to brighter magnitudes, the
galaxies undetected due to their position beneath the horizontal
observed equivalent width limit in Figure \ref{fig:trumpetSch} become a larger
fraction of the overall population, and so the detection fraction
again decreases, and $\alpha$ steepens. The shallowest $\alpha$
therefore, should correspond to the point where
the NB detection limit, $m_{5\sigma}$, is equal to ${m_{\rm{faint}}}$.

\subsection{The Effect of $\alpha$ on $\rho_{\rm{SFR}}$}
\label{alphaSFRD}
The effect of $\alpha$ on $\rho_{\rm{SFR}}$ is dependent on the value of
$\alpha$ itself. For shallow $\alpha$, the effect is small, however,
for steeper slopes the $\rho_{\rm{SFR}}$ estimate is altered more dramatically for
a small change in $\alpha$. For instance, taking the well-constrained
NB921 [O{\sc~ii}] LF at $z=1.46$ and altering $\alpha = -0.91$ by
$\Delta\alpha = 0.1$ 
in either direction (holding $\phi^*$, $L^*$ constant) causes the $\rho_{\rm{SFR}}$ to rise by $\sim 5$ per cent for
steeper (more negative) $\alpha$ but only decrease by $\sim 3$ per cent for
the corresponding $\Delta\alpha = 0.1$ towards shallower (less negative) $\alpha$. Taking the same
approach with the NB921 H$\alpha$ LF at $z=0.4$, $\alpha = -1.14$\,,
and the equivalent $\Delta\alpha = 0.1$ changes in $\alpha$ result in a 11 per cent increase in $\rho_{\rm{SFR}}$
for steeper $\alpha$ and a 8 per cent decrease for a shallower
slope. This behaviour is reflected in the lower panel of Figure
\ref{fig:EWonalpha}, however it is also apparent that $\rho_{\rm{SFR}}$ increases
more rapidly towards smaller limiting EW values for ${m_{\rm{faint}}}$. We
interpret this as the result of an unrealistic limit to ${m_{\rm{faint}}}$
causing the detection fraction of
galaxies to become very small and produce an unrealistic
correction inconsistent with the data.

\subsection{Comparison to Other Studies}
\label{comp}
The NB921--selected [O{\sc~ii}] LF and $\rho_{\rm{SFR}}$  at $z=1.46$ is in excellent
agreement with the results of \cite{Sobral12_double}'s
UDS+Cosmic Evolution Survey (COSMOS)
selection. Conversely though,
comparison of our NB921-selected sample of H$\alpha$ emitters to the
selection of \cite{Sobral12} at $z=0.4$ reveals our
$\rho_{\rm{SFR}}$ is much lower than the UDS+COSMOS
value. However, \cite{Sobral12} found a difference in $\rho_{\rm{SFR}}$ between
their UDS and COSMOS--selected samples, with the UDS value being a
factor of 3 lower
than that for COSMOS. Their best-fitting LF using UDS emitters alone
produces $\rho_{\rm{SFR}}=0.014\; {\rm{M_{\odot}
yr^{-1}}}$ (D. Sobral, private communication) a value slightly higher than
our own. We are able to fully account for this difference if we use the
same method of redshift slice assignment as in \cite{Sobral12}
(where objects with \emph{K} mag $>$ 23.0 were assigned according to \emph{B}$-$\emph{R},
\emph{i}$-$\emph{K} colour). Our work uses a later UDS data release
and so deeper near-infrared imaging results in more accurate
photometric redshifts for faint objects --- these are in good agreement
with \cite{Grutzbauch11}. We also confirm the reliability of our
aperture corrections through the agreement of $\rho_{\rm{SFR}}$ at
$z=0.4$ between \cite{Sobral12}'s UDS-only sample and our H$\alpha$
emitters according to \emph{B}$-$\emph{R},
\emph{i}$-$\emph{K} colour ($\rho_{\rm{SFR}}=0.0138\; {\rm{M_{\odot}
yr^{-1}}}$). The recovery of this $\rho_{\rm{SFR}}$ despite the
differing aperture sizes used for detection (\citealt{Sobral12} use a 3
arcsecond aperture at low redshift) lends support to the
success of our global corrections for aperture
flux loss.

As discussed in Section
    \ref{cos_var} the uncertainty in our results due to the finite
   number of objects detected per redshift slice is
    investigated via a jackknife analysis. However, it is not possible
    to tell if the values of $\rho_{\rm{SFR}}$ we compute are
    representative of the overall $\rho_{\rm{SFR}}$ at each
    redshift. This depends on the large scale structure in
    the UDS at the redshifts we sample. It is not possible to tell
    whether these regions are particularly under- or over-dense without
    a comparison to many more similar surveys. For example, in the
    lowest redshift slice that we sample, [O{\sc~iii}] at $z=0.14$, we
  would require a contiguous volume $\sim 140$ times larger in order for the 
  $1 \sigma$ uncertainty in number counts of galaxies to reduce to
  $20$ per cent according to the calculations of
  \cite{Somerville04}. The issue can be more efficiently addressed by
  observing widely separated regions of sky and comparing the
  variations seen to between such fields to predictions. For example the divergence of $\rho_{\rm{SFR}}$ by a factor of
  $\sim 3$ between this work and
\cite{Sobral12_double}'s UDS+COSMOS measurement at low redshift (H$\alpha$ at $z=0.40$)
is quantitatively consistent with the predictions of of
\cite{Somerville04}. Finally we note that our value of
  $\rho_{\rm{SFR}}$ at $z=1.46$ is in excellent agreement with both
  the $\rho_{\rm{SFR}}$ of \cite{Sobral12_double}'s UDS+COSMOS
  measurement, and \cite{Ly07}'s Subaru Deep Field (SDF)
  measurement. Based on these comparisons we note that our
estimates of $\rho_{\rm{SFR}}$ in the lower redshift slices may differ
from the true values by up to a factor
of $\sim 8$ in
the lowest redshift slice, $z=0.14$. The resultant steep value of
$\gamma$ should therefore not be a cause for concern when the
uncertainty due to cosmic variance is so large.

\section{Conclusions}
\label{conclusions}

We have developed a robust method for evaluating luminosity functions
which correctly handles the selection criteria of NB samples, negating
the need for retrospective corrections to the data set. As a result,
the method is suitable for highly irregular filters, and is
applicable to a broad range of data sets.
This is the first maximum likelihood analysis of
narrow-band emitters using one of the largest
NB-selected data sets to date. Our results can be broadly summarized as follows:

\begin{itemize}
\renewcommand{\labelitemi}{$\bullet$}
\item We have presented maximum-likelihood luminosity functions in 12
redshift slices out to $z=1.64$, and demonstrated that this technique
produces results in concordance with literature luminosity functions
using similar data sets over similar redshift ranges. In particular we
confirm the results of \cite{Sobral12} and \cite{Sobral12_double} in the $z=0.4$ and
$z=1.46$ redshift slices for the H$\alpha$ and [O{\sc~ii}] NB921 samples
respectively.
\item We find the evolution of the $\rho_{\rm{SFR}}$ to rise according
  to $\rho_{\rm{SFR}}$ $\propto$ (1+z)$^{4.58}$ but conclude
  that this is strongly affected by cosmic variance in the small
  volumes probed at low redshift through this technique. 
\item While the areal coverage of this survey is not large enough to place
  constraints on $\rho_{\rm{SFR}}$ at low redshift, the sensitivity to
  very low SFRs that deep narrow-band observations
  provide makes this technique ideal for constraining the faint end
  slope. 
\item We discover that the detection fraction of emission lines plays
  an important role in the determination of the faint end slope. We
  propose that the limit of integration for the total number of
  galaxies capable of hosting an emission line should be
  empirically motivated, and here we base this on the line of interest's maximum observed
  equivalent width in the local Universe.
\item The deep NB data will still provide useful information
  on the population of objects making up the faint end slope, and
  allow an evaluation of SFR with stellar mass. We will present
  this analysis in a future paper.

\end{itemize}

\section*{Acknowledgements}
The authors wish to thank the anonymous referee for detailed
comments which have greatly improved the clarity of this paper. ABD wishes to acknowledge an STFC studentship. JSD acknowledges the
support of the European Research Council through an Advanced Grant, and the support of the Royal Society via a Wolfson Research Merit award.
RJM acknowledges the support of
the European Research Council via the award of a  Consolidator Grant,
and the support of the Leverhulme Trust via the award of a Philip
Leverhulme research prize. The authors wish to acknowledge David Sobral for extended useful discussions,
and Ruth Gr\"{u}tzbauch for the provision of her photometric redshift catalogue.

\bibliographystyle{mn2e}
\bibliography{PAPERS}

\label{lastpage}

\end{document}